\documentclass{jfm}
\usepackage[JFM]{styswitch}
\shorttitle{On the K\'arm\'an-Pohlhausen and Blasius solutions}
\shortauthor{J. Majdalani and L.J. Xuan}

\title{{On the  K\'arm\'an momentum-integral approach and the Pohlhausen paradox}}
\author{Joseph Majdalani\aff{1}\corresp{\email{joe.majdalani@auburn.edu}}
	\and
	Li-Jun Xuan 
}

\affiliation{\aff{1}Auburn University, Auburn, AL 36849, USA
}

\graphicspath{{figures/}}

\usepackage{enumitem, wrapfig} 
\usepackage{upgreek} 
\usepackage[varg]{txfonts}
\usepackage{listings}
\usepackage{bm} 

\newcommand{\dd }{\mathop{}\!\mathrm{d}}

\newcommand{\sbar}{\bar s}
\newcommand{\Fbar}{\bar F}

\usepackage{lettrine}
\usepackage{booktabs}
\usepackage{longtable}
\usepackage{xcolor}
\usepackage{subfig}

\begin{document}
\maketitle

\begin{abstract}
	This work explores simple relations that follow from the momentum-integral equation absent a pressure gradient. The resulting expressions enable us to relate the boundary-layer characteristics of a velocity profile, $u(y)$, to an assumed flow function and its wall derivative relative to the wall-normal coordinate, $y$.  Consequently, disturbance, displacement, and momentum thicknesses, as well as skin friction and drag coefficients, which are typically evaluated and tabulated in classical monographs, can be readily determined for a given profile, $F(\xi)=u/U$. Here $\xi=y/\delta$ denotes the boundary-layer coordinate. These expressions are then employed to provide a rational explanation for the 1921 Pohlhausen polynomial paradox, namely, the reason why a quartic representation of the velocity leads to less accurate predictions of the disturbance, displacement, and momentum thicknesses than using cubic or quadratic polynomials. Not only do we identify the factors underlying this behaviour, we proceed to outline a procedure to overcome its manifestation at any order. This enables us to derive optimal piecewise approximations that do not suffer from the particular limitations affecting Pohlhausen's $F(\xi)=2\xi-2\xi^{3}+\xi^{4}$. For example, our alternative profile, $F(\xi)=(5\xi-3\xi^{3}+\xi^{4})/3$, leads to an order-of-magnitude improvement in precision when incorporated into the  K\'arm\'an-Pohlhausen approach in both viscous and thermal analyses.  Then noting the significance of the Blasius constant, ${\sbar}\approx1.630398$, this approach is extended to construct a set of uniformly valid solutions, including $F(\xi)=1-\exp[-{\sbar}\xi(1+{\tfrac{1}{2}}{\sbar}\xi+\xi^{2})]$, which continues to hold beyond the boundary-layer edge as $y\rightarrow\infty$. Given its substantially reduced error, the latter is shown, through comparisons to other models, to be practically equivalent to the Blasius solution.

\end{abstract}
\section{Introduction}
Following Prandtl's pioneering treatment of boundary layers \citep{Prandtl1904}, and the shape-preserving similarity solution reported by \citet{Blasius1908}, the  K\'arm\'an-Pohlhausen (KP) momentum-integral approach, introduced through two sequential papers by \citet{Karman1921} and \citet{Pohlhausen1921}, is widely regarded as the staple in boundary-layer analysis. This is mainly due to its effectiveness at providing a wealth of useful detail for boundary layers in both low and high Reynolds number flows. Despite its simplicity, the KP approach continues to play a pivotal role in several monographs that devote themselves to boundary-layer theory. These include, for example, those by \citet{Rosenhead1963}, \citet{Schlichting1979}, \citet{Cebeci1998}, \citet{Oleinik1999}, \citet{White2006}, \citet{Schetz2011}, \citet{Fox2015}, \citet{Schlichting2017}, and others. It must be noted, however, that the KP approach is restricted to the analysis of the so-called inner region, with the outer far field referring to the inviscid freestream. 

Given its ease of implementation and broad-brush bookkeeping outcomes in a variety of viscous-flow problems, the KP approach has been applied to a plethora of thermofluid flow configurations.  Examples abound and one may cite studies that involve extensions to Ostwald-de Waele power-law fluids \citep{Bizzell1962}, rarefied gases over porous plates \citep{Jain1972}, rotating three-dimensional boundary layers on cones \citep{Bloor1977}, thermal boundary layers on short rotating blades \citep{Mohanty1977}, forced convection from surfaces immersed in non-Newtonian fluids using planar \citep{Nakayama1986}, axisymmetric \citep{Shenoy1986}, and irregular geometry \citep{Nakayama1988}, forced convection in highly pseudoplastic fluids \citep{Andersson1988}, natural convection over bodies immersed in fluid-saturated porous media \citep{Nakayama1989}, non-Newtonian motions within slider bearings \citep{Bujurke1992},  natural convection over bodies embedded in porous media \citep{Mehta1994}, non-Newtonian boundary layers for low-Reynolds number mud flows \citep{Huang1998}, radial film flows \citep{Rao1998}, asymmetrically-mixed convective motions around horizontal cylinders \citep{Amaouche2003}, heat transfer from infinite isothermal cylinders with both Newtonian and non-Newtonian fluids \citep{Khan2005,Khan2006}, momentum and thermal analyses of power-law fluids over flat plates \citep{Meyers2010}, and forced convection in power-law fluid-saturated porous media \citep{Thayalan2013}. 
 
\begin{figure}
	\centering
	\includegraphics[width=0.67\textwidth]{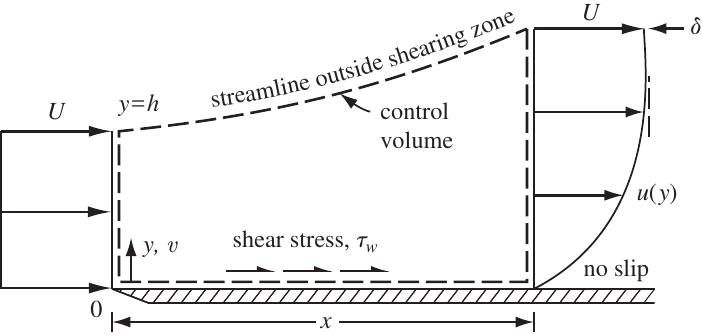}
	\caption{Control volume used in conjunction with the  K\'arm\'an-Pohlhausen momentum-integral analysis for flow over a flat plate.}
	\label{Fig_CV-flat-plate}
\end{figure}

In the present investigation, however, the object is not to extend the KP approach to a new flow configuration. Rather, it stands in reducing the KP formulation in the absence of a pressure gradient to simple relations that display direct connections between its predictions and their true values for flat-plate motion.  Our next objective is to revisit the assumptions made by \citet{Pohlhausen1921}, which have been used at the basis of constructing closed-form analytic expressions for a series of conjectured inner velocity profiles; the latter may be separated into two principal categories: those seeking piecewise approximations, valid only within the boundary layer \citep{Pohlhausen1921,Schlichting1955}, and those pursuing monotonically-increasing continuous profiles that merge smoothly into the far field as $y\rightarrow \infty$ \citep{Blasius1908,Bairstow1925,Parlange1981,Boyd1997,Liao1997,Liao1999a,Liao1999b,Boyd2008, Iacono2015}.  We begin by considering those that extend over $0\le y\le \delta $, where $y=\delta $ represents the edge of the viscous layer and $y$ denotes the wall-normal coordinate. As usual, we assign $U$ to the freestream velocity and provide a labelled sketch of the streamline-bounded control volume and coordinate system in \figref{Fig_CV-flat-plate}.

Since the majority of continuous flow formulations aim at producing uniformly valid analytical solutions to the Blasius equation, we find it necessary to evaluate the viability of the corresponding boundary conditions, which are embedded in a KP formulation, vis-\`a-vis those predicted by a differential technique, to retrieve comparable velocity approximations.  This two-pronged effort is carried out while taking into account the subtle connection between the Blasius similarity variable, $\eta$, and the normalized coordinate, $\xi$, which arises in the integral formulation. By closely examining parallel predictions from the integral and differential approaches for the same flow configuration, we are able to identify previously-used boundary conditions which, when replaced by more suitable constraints, lead to substantially more accurate flow profiles.  More specifically, we show that Pohlhausen's quartic polynomial, $u/U=F(\xi )=2\xi -2\xi ^{3} +\xi ^{4} $, which has been routinely used in viscous and thermal analyses, with and without pressure gradients, can be definitively amended by allowing it to satisfy a more realistic boundary condition at the edge of the boundary layer. By so doing, we arrive at an alternative representation, $F(\xi )=(5\xi -3\xi ^{3} +  \xi ^{4} )/3$, which will be shown to outperform other piecewise approximations in its class while predicting both viscous and thermal layer characteristics.  It will also open several avenues of research inquiry, including higher-order extensions.
  
 Recognizing the utility of this approach in deriving spatially-restricted piecewise approximations, and since no flat-plate study is complete without referral to the Blasius equation, we circle back and apply the same procedure to obtain a continuous approximation to the Blasius problem over its semi-infinite domain. The latter continues to serve as an important benchmark and testing ground for new viscous and heat transfer approximations as well as stability and transition analyses entailing both traditional and non-Newtonian fluids, with some including wall roughness, inclination, or suction \citep[see, for example,][]{Levin2005,Zhao2017,Depaula2017,Ng2017,Hack2017,Tsigklifis2017,Hewitt2018,Tsang2018,Tsang2019,Brynjell2019,Beneitez2019}.  
 
 By taking advantage of the Blasius constant at the wall,  $F'(0)=\sbar\approx 1.630398$, and the crucial role that it plays in prescribing the trajectory of the solution under construction, we identify five essential boundary conditions that are inspired by Pohlhausen's classic requirements; the latter are sequentially applied in a manner to produce a compact, decaying exponential function that entails a maximum $L_{2}$ error of $6.4\times 10^{-4}$.  Such a small discrepancy places it within the same margin of error affecting the Blasius equation itself. We close by extending this approach to higher orders and by comparing the one-term expression, $F(\xi )=1-\exp [-\sbar\xi (1+{\tfrac{1}{2}} \sbar\xi +\xi ^{2} )]$, to other available profiles in the semi-infinite range of $0\le \xi <\infty $. Our verification tools and error analyses are systematically used to demonstrate that the decaying exponential formulation matches the Blasius solution virtually identically, not only from the wall to the edge of the boundary layer, but all the way to infinity. 
 
From an organizational standpoint, the paper is divided into three main sections, besides the introduction and conclusion. In \cref{Reduced momentum-integral relations}, several reduced momentum-integral relations are presented to help connect the fundamental boundary-layer properties, including the wall-normal velocity and vorticity, directly to an assumed velocity profile. This is followed in \cref{Improved effectiveness of the KP integral approach} with a procedure to rapidly evaluate boundary-layer properties using piecewise velocity approximations. Along the way, the paradoxical behaviour of Pohlhausen's high-order polynomial approximations is brought to light and discussed.  The specific factors leading to unexpected deviations from the correct Blasius predictions with increasing polynomial orders are also identified. By introducing an alternative set of boundary conditions, the ability to construct high-order polynomials with increasing degrees of accuracy is demonstrated, starting with a quartic polynomial.  The latter is shown to reduce the overall $L_2$ error associated with the KP method by one order of magnitude. Next, the improved effectiveness of the KP approach, when used in concert with more realistic velocity and temperature profiles, is confirmed in the context of thermal analysis. More specifically, in the case of flat-plate heat transfer, it is shown that the KP approach is capable of predicting the local Nusselt number within a half percent relative error. Finally, the same tools developed in \cref{Reduced momentum-integral relations,Improved effectiveness of the KP integral approach} are extended in \cref{Development of a continuous solution for the Blasius problem} to the systematic development of continuous, decaying exponential profiles that mirror the characteristic behaviour of the Blasius solution almost flawlessly. The accuracy of these simple approximations relative to other comparable models in the literature is also examined and discussed.
 
\section{Reduced momentum-integral relations}\label{Reduced momentum-integral relations}
In this section, the geometry and KP formulation are presented in the context of flat-plate analysis in conjunction with the use of a conjectured viscous-flow approximation for $u(y)$. Direct relations that link $u(y)$, or more generally, $F(\xi)$, to traditionally computed properties, such as the disturbance, displacement, and momentum thicknesses, are also established. These will play a key role in improving the effectiveness of the KP integral procedure, starting with the classical flat-plate flow problem that we revisit here as a test bed for analysis.

\subsection{Flat-plate configuration}
 \noindent Our schematic of the physical setting in \figref{Fig_CV-flat-plate} sets the sharp edge of the plate at $(x,y)=(0,0)$ and shows the fluid shearing at the wall as it sweeps along the plate, where a tangential stress, $\tau_w$, is exerted. The velocity distribution, $u(y)$, at any downstream station, $x$, exhibits a smooth drop-off before vanishing at the wall. The normal `upwelling' velocity is taken to be $v$ and the disturbance or shear-layer thickness is labelled as $\delta $. Streamlines situated outside this layer are deflected vertically by a distance $\delta\raisebox{+.15ex}{*}$, which defines the displacement thickness. The upper streamline slings upwardly from $y=h$ at $x=0$ to $y=\delta =h+\delta\raisebox{+.15ex}{*}$ at $x=L,$ in a control volume of length $L$. In this work, we assume that the freestream just upstream of the sharp edge is uniform with $u=U=\text{const}$. It is useful to normalize the velocity and transverse coordinate using $F(\xi )\equiv u/U$ and $\xi\equiv y/\delta$.
%
%

\subsection{Reduced momentum-integral relations}

\noindent By focusing on the incompressible planar motion over a flat plate at zero incidence, one can use continuity to determine $h=\int_{0}^{\delta}(u/U)\dd y=\delta \int_{0}^{1}F(\xi) \dd{\xi}$. 
Then using $\rho$ for density, a momentum balance enables us to write, as detailed by \citet{Schlichting1979} or \citet{White2006}:
\begin{equation} \label{Eq_unsteady_momentum_integral} 
\frac{C_{f} }{2} \equiv \frac{\tau _{w} }{\rho U^{2} } =\frac{1}{U^{2} } \frac{\partial }{\partial t} \left(U\delta\raisebox{+.15ex}{*}\right)+\frac{\partial \theta }{\partial x} +\left(2\theta +\delta\raisebox{+.15ex}{*}\right)\frac{1}{U} \frac{\partial U}{\partial x} 
\end{equation} 
where $ C_f $ stands for the skin friction coefficient. As usual, the incompressible displacement and momentum thicknesses can be expressed as
\begin{equation} \label{Eq_displacement_thickness}
\delta\raisebox{+.15ex}{*}= \int_{0}^{\infty}{\frac{U-u}{U}\dd y} = \delta \int_{0}^{\infty}(1-F) \dd \xi \approx \delta \int_{0}^{1}(1-F) \dd \xi
\end{equation} 
\begin{equation} \label{Eq_momentum_thickness}
 \theta= \int_{0}^{\infty}{ \frac{u}{U}\frac{U-u}{U}\dd y} = \delta \int_{0}^{\infty} F(1-F) \dd \xi \approx \delta \int_{0}^{1}F(1-F) \dd \xi
\end{equation} 
We recall that the semi-infinite to finite domain approximations are permitted because of the negligible contributions of the integrands for $y>\delta $. In fact, these approximations become strict identities for piecewise velocity profiles that remain equal to unity for $y\geq\delta$.  Moreover, these expressions hold true for any incompressible fluid, be it laminar or turbulent, with constant or variable pressure.  
Next, we find it useful to define the non-dimensional displacement and momentum thicknesses, $\alpha$ and $\beta$, along with the shape factor, $H$, using
\begin{equation} \label{Eq_nond_displacement_thicknesses}
\alpha\equiv \frac{\delta\raisebox{+.15ex}{*}}{\delta}\approx  {\tilde \alpha} \equiv  \int_{0}^{1}(1-F) \dd \xi   \quad\quad \beta \equiv \frac{\theta}{\delta}\approx {\tilde \beta}  \equiv \int_{0}^{1}F(1-F)\dd \xi     \quad \text{ and}\quad H \equiv \frac{\delta\raisebox{+.15ex}{*}}{\theta}=\frac{\alpha}{\beta}
\end{equation} 
where ${\tilde \alpha}$ and ${\tilde \beta}$ represent the approximated integral values of $\alpha$ and $\beta$ over the finite $[0,1]$ interval.  Note the strict equalities of ${\tilde \alpha}=\alpha$ and ${\tilde \beta}=\beta$ for piecewise analytic profiles exhibiting the property $F(\xi \geq 1)=1$.  For this reason, the distinction of using ${\tilde \alpha}$ or ${\tilde \beta}$ is only necessary in the calculation of integral properties associated with uniformly valid, continuously analytic profiles for which the default integrals are evaluated over a semi-infinite domain.  It may be also instructive to note that $\alpha$, $\beta$ and $H$ are pure constants that depend strictly on the assumed velocity distribution $F$. \footnote{For uniformly valid, non-piecewise velocity profiles that continue to increase in the far field for $\xi>1$, restoring the semi-infinite bounds on $\alpha$ and $\beta$ in \cref{Eq_displacement_thickness,Eq_momentum_thickness} is necessary to avoid underestimating their values.  In the case of the Blasius solution, for example, the semi-infinite bounds will result in $0.284\%$ and $0.733\%$ higher values for $\alpha$ and $\beta$, respectively.}
Since the focus of this study is on the steady form of \eqref{Eq_unsteady_momentum_integral}, we continue with 
\begin{equation}
	\dfrac{C_{f} }{2} =\dfrac{\dd \theta }{\dd x} +\left(2+H\right)\dfrac{\theta }{U} \dfrac{\dd U}{\dd x} \quad      \text{ or }    \quad \dfrac{C_{f} }{2\beta} =\dfrac{\dd\delta }{\dd x} +\left(2+H\right)\dfrac{\delta }{U} \dfrac{\dd U}{\dd x}  
	\label{Eq_steady_momentum_integral}
\end{equation}
which, for a spatially-invariant freestream, becomes
\begin{equation} 
	C_{f}=\dfrac{2\tau _{w} }{\rho U^{2} } = 2\beta\dfrac{\dd\delta }{\dd x} 
	\label{Eq_steady_uniform_momentum_integral}
\end{equation}

\subsection{Direct connections to a conjectured velocity profile}
\label{Direct connection to an assumed velocity profile}
\noindent To make further headway, the following procedure may be used. First, a piecewise analytic velocity may be posited in the form
 \begin{equation}
	\dfrac{u}{U} =\left\{\begin{array}{l} {F(\xi ) \quad 0\le \xi \le 1} \\ 
	{1\quad\; \quad 1<\xi <\infty } \end{array}\right.     
\end{equation}
Once $F(\xi )$ is specified, the constants $\alpha$ and $\beta$ may be readily determined from their simple integrals in \eqref{Eq_nond_displacement_thicknesses}. Then using the stress-strain relation at the wall, $\tau_w$ can be written as a function of the disturbance thickness and the gradient of the assumed velocity at $y=\xi=0$. On the one hand, we recover
 \begin{equation}\label{Eq_wall_shear_stress1}
 \tau_w=\mu \left.\frac{\dd u}{\dd y}\right|_{y=0}=\frac{\mu U}{\delta}s 
 \quad\quad \text{where}\quad\quad s\equiv\left.\frac{\dd F}{\dd \xi}\right|_{\xi=0}=F'(0)
 \end{equation}
 where $\mu$ is the dynamic viscosity and primes indicate differentiation with respect to $\xi$. On the other hand, $\tau_w$ may be related to the axial gradient of the disturbance thickness based on \eqref{Eq_steady_uniform_momentum_integral}:
 \begin{equation}\label{Eq_wall_shear_stress2}
 \tau _{w} = \rho U^{2} \beta \dfrac{\dd\delta }{\dd x} 
 \end{equation}
These two expressions may be equated to obtain 
 \begin{equation}\label{Eq_ODE}
 \delta\dd \delta=\frac{\nu s}{U\beta}\dd x 
 \end{equation}
where $\nu$ is the kinematic viscosity. For a boundary layer that starts to develop at $x=0$, one can use $ \delta(0)=0$ and integrate \cref{Eq_ODE} from the plate's leading edge to any station $x$. After some rearranging, one recovers the traditional dependence on the local Reynolds number, namely,  
 \begin{equation}\label{Eq_a_constant} 
 \frac{\delta}{x}=\frac{a}{\sqrt{Re_x}}\quad {\rm and}\quad \dfrac{\dd\delta }{\dd x} =\dfrac{(a/2)}{\sqrt{Re_{x} } };  \quad a \equiv \sqrt{\frac{2s}{\beta}}
 \end{equation}
where $Re_x=Ux/\nu$.  Note the clear connection between the traditional constant $a$ and both the non-dimensional momentum thickness, $\beta$, and the initial slope, $s=F'(0)$.  For the entire class of polynomial approximations of the form $F=a_0+a_1\xi +a_2\xi^2+\dots$, $s$ is simply $a_1$, which can be found by inspection. Another substitution into the friction coefficient in \eqref{Eq_steady_uniform_momentum_integral} leads to
 \begin{equation}\label{Eq_Cf_constant}
C_f=\frac{2\nu s}{U\delta}=\frac{b}{\sqrt{Re_x}}; \quad b \equiv \sqrt{2\beta s} 
 \end{equation}
 The remaining boundary-layer properties follow suit. Recalling that the displacement and momentum thicknesses are deducible from their reversed definitions, $\delta\raisebox{+.15ex}{*}=\delta \alpha$ and $\theta =\delta \beta$, we get
  \begin{equation}\label{Eq_displacement_momentum_thicknesses}
 	\dfrac{\delta\raisebox{+.15ex}{*}}{x} =\dfrac{c}{\sqrt{Re_{x} } };\quad c\equiv a\alpha \quad {\rm and} \quad \dfrac{\theta}{x} =\dfrac{d}{\sqrt{Re_{x} }} ;\quad d\equiv a\beta=b
 \end{equation}
This outcome confirms the equivalence of $\theta/x$ and the skin friction coefficient in \eqref{Eq_Cf_constant}.  
 Interestingly, we find the products and ratios of the traditional constants $ b $ and $ a $, which are often tabulated in various monographs \citep{Fox2004,White2006,Fox2015}, to be proportional to $s$ and $\beta$:
\begin{equation}\label{Eq_ab2s}
ab=ad=2s \quad {\rm and} \quad \frac{b}{a}=\frac{d}{a}=\beta
\end{equation}
It follows that
\begin{equation}\label{Eq_ab_product} 
\frac{\delta}{x} C_f = \frac{\delta \theta}{x^2} = \frac{2 s}{Re_x}    \quad {\rm and} \quad  \frac{x}{\delta} C_f  =\beta
\end{equation}
Practically, for a given profile $F$, it is only necessary to calculate $\alpha$, $\beta$ and $s$ to be able to deduce $a=\sqrt{2s/\beta}$, $b=d=a\beta$ and $c=a\alpha$.  As for the global drag coefficient, it may be evaluated using an elegant expression introduced by \citet{Karman1921},
\begin{equation}
	\label{Eq_drag}
	C_{D} =\frac{1}{L} \int _{0}^{L}C_{f} (x)\dd x =\frac{1}{L} \int _{0}^{L}\left(2\frac{\dd\theta }{\dd x} \right)\dd x =\frac{2}{L} \theta (L) =\frac{2a\beta}{\sqrt{Re_{L} } } =\frac{2b}{\sqrt{Re_{L} } } 
\end{equation}
In view of \cref{Eq_Cf_constant}, it is clear that $C_{D} = 2 C_f$.

  At this point, we can revert back to the continuity equation for the purpose of retrieving valuable estimates for $v$ and its maximum value at the boundary-layer edge.  We start by switching $u(y)$ to $F(\xi)$ in the continuity integral,
\begin{equation}\label{Eq_v_velocity_derivation1}
\begin{split}
v=&-\frac{\partial }{\partial x} \int _{0}^{y}u \dd y =-\frac{\partial }{\partial x} \int _{0}^{y}UF(\xi) \dd y
=-U\int _{0}^{y}\frac{\dd F(\xi)}{\dd \xi}\frac{\dd \xi}{\dd x} \dd y\\
\end{split}
\end{equation}
This enables us to identify the presence of $\delta(x)$, which may be readily expanded and rearranged within the integrand using
\begin{equation}\label{Eq_v_velocity_derivation2}
\begin{split}
v=&-U\int _{0}^{y}\frac{\dd F(\xi)}{\dd \xi}\frac{\dd [y/\delta(x)]}{\dd x} \dd y
=U\int _{0}^{y}\dfrac{y}{\delta^2}F^{\prime}(\xi)\frac{\dd \delta(x)}{\dd x} \dd y
=U\int _{0}^{\xi}\xi F^{\prime}(\xi)\frac{\dd \delta(x)}{\dd x} \dd \xi\\
\end{split}
\end{equation}
Moreover, by virtue of \cref{Eq_a_constant}, the axial gradient of $\delta(x)$ can be eliminated in favour of the local Reynolds number. The resulting expression can be further normalized by $U$ and integrated by parts.  One gets:
 \begin{equation}\label{Eq_v_velocity_derivation3}
\begin{split}
\frac{v}{U}=&\frac{a }{2\sqrt{Re_x}}\int _{0}^{\xi}\xi F^{\prime}(\xi) \dd \xi
=\frac{a }{2\sqrt{Re_x}}\left[\xi F(\xi) -  \int _{0}^{\xi} F(\xi) \dd \xi \right]= \frac{G(\xi)}{\sqrt{Re_{x} } } \\
\end{split}
\end{equation}
where the normal velocity's characteristic function $G(\xi)$ can be determined directly from
\begin{equation}\label{Eq_v_velocity_max}
G(\xi)= \frac{a}{2}\int _{0}^{\xi}\xi F^{\prime}(\xi) \dd \xi=\frac{a}{2}\left[\xi F(\xi) -  \int _{0}^{\xi} F(\xi) \dd \xi \right]
\end{equation}
Naturally, $G(\xi)$ will depend on the profile at hand.  For piecewise velocity representations with $F(\xi \geq 1)=1$, the peak normal velocity may be shown to occur at $y=\delta$ where
\begin{equation}
\label{Eq_v_velocity_solution}
\left(\frac{v}{U} \right)_{\delta } \equiv \left. \frac{v}{U} \right|_{\xi=1 } =\frac{G_{\delta}}{\sqrt{Re_{x} } };\quad\quad G_{\delta} \equiv G(1)  
\end{equation} 
To determine $G_{\delta}$ explicitly, it is helpful to recreate the definition of ${\tilde \alpha}$ in \cref{Eq_v_velocity_max}. By simple manipulation, we arrive at

\begin{equation}\label{Eq_v_max_coefficient}
G_{\delta}=\frac{a}{2}\left[F(1) -  \int _{0}^{1} F(\xi) \dd \xi \right]=\frac{a}{2}\left\{ F(1) -1 +  \int _{0}^{1} \left[ 1- F(\xi)\right]  \dd \xi \right\}
=\frac{a}{2}\left[F(1) - 1+{\tilde \alpha} \right]
\end{equation}
This expression is somewhat illuminating. It shows that for all assumed velocity approximations that observe the $F(\xi\geq 1)=1$ condition, the maximum normal velocity coefficient reduces to $G_{\delta}=a{\tilde \alpha}/2=a\alpha/2=c/2$ by way of \cref{Eq_displacement_momentum_thicknesses}. As such, and in reference to standard skin friction and momentum thickness tables \citep{Schlichting1979,Fox2004,White2006,Fox2015}, the peak normal velocity coefficients, ${\sqrt{Re_{x} }}(v/U)_{\rm max}$, are simply half of the tabulated values for ${\delta\raisebox{+.15ex}{*}}{\sqrt{Re_{x} }/{x}}$.  

However, for those approximations that observe Prandtl's $F(1)=0.99$ cutoff requirement on $u/U$, we get $G_{\delta}=a{\tilde \alpha}/2-0.005a$. Moreover, for uniformly valid velocity profiles that continue to increase past $\xi=1$, the maximum normal velocity occurs at infinity rather than the edge of the disturbance thickness.  Besides the normal velocity at $\xi=1$, a slightly larger value may be realized at
\begin{equation}
\label{Eq_v_velocity_solution_inf}
\left(\frac{v}{U} \right)_{\rm max } \equiv \left. \frac{v}{U} \right|_{\xi\rightarrow \infty } =\frac{G_{\infty}}{\sqrt{Re_{x} } };\quad\quad G_{\infty} \equiv G(\infty)  
\end{equation} 
Here too, the maximum $G_{\infty}$ coefficient may be readily deduced from \cref{Eq_v_velocity_max}. Making use of $F(\infty)=1$, we can add and subtract the variable $\xi$ by putting
\begin{equation}\label{Eq_v_max_coefficient_inf}
G_{\infty}=\frac{a}{2}\left[\xi F(\xi) -  \int _{0}^{\xi} F(\xi) \dd \xi \right]_{\xi\rightarrow \infty } =\frac{a}{2}\left\{    \xi \left[F(\xi) -1\right]  +  \int _{0}^{\xi} \left[1-F(\xi)\right] \dd \xi \right\}_{\xi\rightarrow \infty } =\frac{a \alpha}{2}
\end{equation}
This simple outcome is, of course, contingent upon $ \xi \left[F(\xi) -1\right] \rightarrow 0$ as $\xi \rightarrow \infty.$  Such will be the case when
\begin{equation}\label{Eq_G_inf}
 \lim_{\xi\rightarrow \infty} {\frac {F(\xi) -1 } {\xi^{-1}} } = \lim_{\xi\rightarrow \infty} {\frac {F'(\xi) } {-\xi^{-2}} }=0 \quad {\rm or } \quad F'(\infty) = o \left( \xi^{-2} \right)
\end{equation} 
The last condition in \cref{Eq_G_inf} is safely secured because, in boundary-layer theory, the gradient of the velocity in the far field is known to approach zero exponentially, i.e., faster than any polynomial order.  Furthermore, the use of $(v/U)_{\rm max}=a\alpha/2$ is consistent with corresponding estimates reported in the literature \citep{Fox2004,White2006}.  Another benefit of these relations is that they enable us to link the key characteristic properties explicitly to $a$, $\alpha$, and $F$.

At this juncture, having determined both $u$ and $v$, we can proceed by evaluating the boundary-layer vorticity, $ \omega $.  Being a cross-derivative function, the outcome is straightforward. We get
\begin{equation}\label{Eq_vorticity}
\omega = \frac{\partial v}{\partial x} - \frac{\partial u}{\partial y}=
-\frac{U}{\delta} \left[ \frac{a}{2Re_x} G'(\xi) + F'(\xi)  \right]  \quad  \text{or}  \quad \omega\raisebox{+.15ex}{*} =\frac{\omega \delta}{U} = - \frac{a}{2Re_x} G'(\xi) - F'(\xi) 
\end{equation}
where $\omega\raisebox{+.15ex}{*}$ denotes the non-dimensional vorticity. The last expression may be further simplified using \cref{Eq_v_velocity_max} into
\begin{equation}\label{Eq_vorticity_nondimensional}
\omega\raisebox{+.15ex}{*} =- F'(\xi)\left(1  + \frac{a^2 \xi}{4 Re_x}   \right)   \quad  \text{and so}  \quad  \left.\omega\raisebox{+.15ex}{*}\right|_{\rm max} =\left.\omega\raisebox{+.15ex}{*}\right|_{\xi=0}=-F'(0)
\end{equation}

Compared to the normal velocity and its peak value in \cref{Eq_v_velocity_derivation3,Eq_v_velocity_solution}, which are inversely proportional to the square root of the Reynolds number, the $G'$-dependent term in \cref{Eq_vorticity} is appreciably smaller, being proportional to $Re_x^{-1}$.  As such, $\omega\raisebox{+.15ex}{*}\approx -F'(\xi)$ is a suitable one-term approximation for the vorticity, whose peak value at the wall, $\omega\raisebox{+.15ex}{*}(0)= -s$, becomes the negative of the so-called `connection parameter' in the corresponding Blasius problem. This equality is consistent with the foundational role that shear-layer vorticity plays in prescribing the behaviour of fluid motion.

Before leaving this section, it may be helpful to reiterate that the procedure just described is fairly straightforward and facilitates the evaluation of most salient boundary-layer features for a given velocity distribution.  Once the defining integrals for $\alpha$ and $\beta$ are determined, and $s$ is in hand, the remaining properties can be calculated with minimal effort using \cref{Eq_a_constant,Eq_Cf_constant,Eq_displacement_momentum_thicknesses,Eq_ab2s,Eq_ab_product,Eq_drag,Eq_v_velocity_derivation1,Eq_v_velocity_derivation2,Eq_v_velocity_derivation3,Eq_v_velocity_max,Eq_v_velocity_solution,Eq_v_max_coefficient,Eq_vorticity,Eq_vorticity_nondimensional}. In fact, the ability to connect various boundary-layer properties directly to an assumed velocity profile will prove essential in the remainder of this work.

\section{Improved effectiveness of the KP integral approach}\label{Improved effectiveness of the KP integral approach}

For decades, integral methods have been dominated by the use of intuitive, guessed velocity profiles. As indicated earlier, the K\'{a}rm\'{a}n--Pohlhausen approach is named as such after two concurrent articles by \citet{Karman1921}, who provided the integral formulation, and \citet{Pohlhausen1921}, who presented a systematic procedure for constructing and integrating suitable velocity approximations. A side benefit of the KP approach is that once a reasonable form of $u$ is conjectured, accurate estimates of the boundary-layer properties can be expected because integration tends to wash out the positive and negative deviations of the assumed profile from the exact velocity distribution. However, this procedure is not without limitations.

\subsection{Direct evaluation of properties using piecewise velocity approximations}\label{Direct evaluation of properties using piecewise velocity approximations}

\noindent To illustrate the utility of the foregoing relations, the boundary-layer properties will be evaluated using several closed-form approximations that have been widely used in the literature.  The accuracy of these profiles will then be gauged relative to the classic Blasius solution for flow over a flat plate.  We begin by examining five piecewise models that are defined for $0\le y\le \delta $, and which are set equal to unity over the semi-infinite domain, $1<y<\infty $. These correspond to Pohlhausen's quadratic, cubic, and quartic profiles \citep{Pohlhausen1921}: 
\begin{equation}\label{Eq_Pohlhausen}
	\dfrac{u}{U} = 2\dfrac{y}{\delta } -\dfrac{y^{2} }{\delta^{2} } ,\quad \dfrac{u}{U} = \dfrac{3}{2} \dfrac{y}{\delta } -\dfrac{1}{2} \dfrac{y^{3} }{\delta ^{3} }, \quad \dfrac{u}{U} = 2\dfrac{y}{\delta } -2\dfrac{y^{3} }{\delta ^{3}} +\dfrac{y^{4} }{\delta ^{4}}    
\end{equation}
Schlichting's sinusoidal profile \citep{Schlichting1955}:
\begin{equation}\label{Eq_Schlichting}
	\dfrac{u}{U} = \sin \left(\dfrac{\pi y}{2\delta } \right) 
\end{equation}
and a rationally-optimized quartic profile: 
\begin{equation}\label{Eq_Majdalani-Xuan}
	\dfrac{u}{U} =\dfrac{5}{3} \dfrac{y}{\delta } -\dfrac{y^{3} }{\delta ^{3} } +\dfrac{1}{3} \dfrac{y^{4} }{\delta ^{4} } 
\end{equation}

Using their normalized forms, these profiles are collected in \tblref{Table_coefficient_comparison} and compared to the classic solution due to Blasius (1908). Note that relative percentage errors are provided below each estimate along with the overall $L_{2} =[\int _{0}^{1}(F-\Fbar)^{2} \dd \xi]^{1/2} $ error across the viscous layer, where a `barred' quantity refers to its Blasius value. This is performed to help quantify the overall deviation in each approximation from the traditionally reported Blasius values \citep{Schlichting1979,Fox2004,White2006,Anderson2017,Fox2015}. Properties that are not shown can be easily deduced because, for laminar flow, $(\theta /x)\sqrt{Re_{x} } =C_{f} \sqrt{Re_{x} } =b$, which is already tabulated, and $C_{D} \sqrt{Re_{L} } $ is twice this constant.  Conversely, the maximum wall-normal velocity, ${\sqrt{Re_{x} }}(v/U)_{\rm max}$, is simply half of the entries for ${\delta\raisebox{+.15ex}{*}}{\sqrt{Re_{x} }/{x}}$, as shown in \cref{Direct connection to an assumed velocity profile}.

Apart from the optimized quartic profile, whose error does not exceed 1.7\% in the $0 \leq \xi \leq 1$ interval, it may be seen that the maximum error in each case varies between 6 and 17\%, which is typical of integral theories. Some profiles, however, appear to be distinctly more precise than others, namely, in mirroring the behaviour of the Blasius solution. What is most perplexing, perhaps, is what some academicians have come to identify, rather informally, as the Pohlhausen paradox: one may notice that as we move from Pohlhausen's second-order to his fourth-order polynomial --which is capable of securing additional boundary conditions,-- the agreement with the Blasius solution does not improve, as one would expect, but rather deteriorates.  Being part of a widely taught subject, this inconsistency has often puzzled fluid mechanics learners and instructors alike. From \tblref{Table_coefficient_comparison}, it can be seen that the errors in estimating the non-dimensional displacement and momentum thicknesses $\{\alpha,\beta\}$ leap by one order of magnitude, i.e., from $\{3.1\%, 0.25\%\}$ to $\{13\%, 12\%\}$, while the overall $L_{2} $ error increases from 0.02 to 0.05, when Pohlhausen's quadratic polynomial is exchanged with its quartic counterpart. The error in predicting the boundary-layer thickness also increases from 9.5\% to a whopping 17\%.

\begin{table}
	\centering
	\begin{center}
		\begin{tabular}{l|c|c|c|c|c|c|c} 
			\toprule
			$ F(\xi)=\dfrac{u}{U}$
			& $\alpha=\dfrac{\delta\raisebox{+.15ex}{*}}{\delta}$& $\beta=\dfrac{\theta}{\delta}$   & $H=\dfrac{\delta\raisebox{+.15ex}{*}}{\theta}$ & $ \dfrac{{\delta }}{x}\sqrt {R{e_x}}  $  & $ C_f\sqrt {R{e_x}} $ & $ \dfrac{{\delta\raisebox{+.15ex}{*}}}{x}\sqrt {R{e_x}} $   & $ L_2$ error\\
			\midrule
			
			$ 2\xi-\xi^2 $ &  $ 0.333 $ & $ 0.133 $    & $ 2.500 $  &	$ 5.477 $ 
			& $ 0.730 $  & $ 1.826 $ & $ 0.020 $\\
			& $ 3.1\%$ & $ 0.25\%$  & $ 3.5\%$  &  $9.5\%$  &  $ 10\%$ &  $6.1\%$\\
			\midrule
			
			$ \tfrac{3}{2}\xi-\tfrac{1}{2}\xi^3 $ &  $ 0.375 $& $ 0.139 $    & $ 2.692 $  &	$ 4.641 $ & $ 0.646 $  & $ 1.740 $  & $ 0.034 $\\
			& $ 9.0\%$& $ 4.7\%$   & $ 4.0\%$  &  $ 7.2\%$  &  $ 2.6\%$ &  $ 1.1\%$\\
			\midrule
			$2\xi-2\xi^3+\xi^4 $   &  $ 0.300 $& $ 0.118 $  & $ 2.554 $  &	$ 5.836 $ & $ 0.685 $  & $ 1.751 $  & $ 0.054 $\\
			& $13\%$& $12\%$   & $ 1.4\%$  &  $17\%$  &  $ 3.2\%$&  $ 1.8\%$\\ 
			\midrule
			
			$\sin\left(\tfrac{1}{2}\pi\xi\right) $  &  $0.363 $& $ 0.137 $   & $ 2.660 $  &	$ 4.795 $ & $ 0.655 $ & $ 1.743 $  & $ 0.021 $\\
			& $ 5.6\%$& $2.7\%$   & $ 2.7\%$  &  $ 4.1\%$  &  $ 1.3\%$&  $ 1.3\%$\\
			\midrule
			
			$\tfrac{5}{3}\xi-\xi^3+\tfrac{1}{3}\xi^4  $  &  $ 0.350 $& $ 0.134 $   & $ 2.618 $  &	$ 4.993 $ & $ 0.668 $ & $ 1.748 $ & $ 0.008 $\\
			& $ 1.7\%$& $ 0.52\%$   & $ 1.1\%$  &  $ 0.13\%$  &  $ 0.53\%$ &  $ 1.6\%$\\
			\midrule
			Blasius (1908)  &  $ 0.344 $& $ 0.133 $    & $ 2.59 $  &	$ 5 $ & $ 0.664 $ & $ 1.72 $ &      \\
			\bottomrule

		\end{tabular}
			\caption{Boundary-layer predictions from five piecewise analytic solutions with their errors relative to the classic Blasius values \citep{Fox2004,White2006,Fox2015}.}
			\label{Table_coefficient_comparison}
	\end{center}	

\end{table}

\subsection{Paradoxical behaviour of Pohlhausen's polynomials}\label{Paradoxical behavior of Pohlhausen's polynomial approximations}

\noindent So one may wonder, what is controlling the accuracy of these profiles?  For several decades, the underlying assumption has been that the fidelity of the conjectured profiles could be improved by securing additional boundary conditions that are representative of the Blasius model. In this vein, \citet{Pohlhausen1921} proposed five cardinal requirements which, apart from the wall adherence and smooth merging with the freestream conditions, i.e., the three constraints satisfied by the quadratic polynomial in \cref{Eq_Pohlhausen}, piled on two further requirements on the shear stress: one at the wall and one at the edge of the viscous domain. To summarize, the first three foundational constraints consist of: 
\begin{align}
u(x,0) &=0 \quad \text{(no slip)} \label{Eq_noslip} \\
u(x,\delta ) &=U  \quad  \text{and}  \quad  \left. \dfrac{\partial u}{\partial y} \right|_{y=\delta } =0 \quad  \text{(smooth merging with the far field)} \label{Eq_smoothedge_BC}
\end{align}
As for the fourth condition, it serves to maintain the correct momentum balance at the wall between the non-vanishing components of the boundary-layer equation via
\begin{equation}\label{Eq_wall_momentum_balance_BC}
	\mu \left. \dfrac{\partial ^{2} u}{\partial y^{2} } \right|_{y=0} =\dfrac{\dd p}{\dd x} \quad   \text{ (momentum balance at $y=0$)}; \quad   \dfrac{\dd p}{\dd x} =-\rho U\dfrac{\dd U}{\dd x}  \text{ (Euler's far field)}
\end{equation}
\noindent Here the pressure gradient within the boundary layer is expressed in terms of the freestream velocity using Euler's equation.  One may verify that not only do Pohlhausen's cubic and quartic polynomials satisfy this momentum balance requirement, but so do Schlichting's sinusoidal and the optimized quartic profiles.  

Lastly, Pohlhausen's fifth condition, i.e., the one that we scrutinize here, assumes zero shear at the edge, which translates into a vanishing curvature at $y=\delta :$
\begin{equation}\label{Eq_edge_shear_BC}
\left. \frac{\partial ^{2} u}{\partial y^{2} } \right|_{y=\delta } =0
\end{equation}
Of the five profiles described so far, only Pohlhausen's quartic polynomial satisfies this condition identically. Incidentally, it also proves to be the most imprecise, as it leads to the largest overall $L_{2} $ error and a 17\% discrepancy in predicting the boundary-layer thickness relative to the traditionally reported Blasius values \citep{Fox2004,White2006}.   

\begin{figure}
	\subfloat[]{\includegraphics[width=0.49\textwidth]{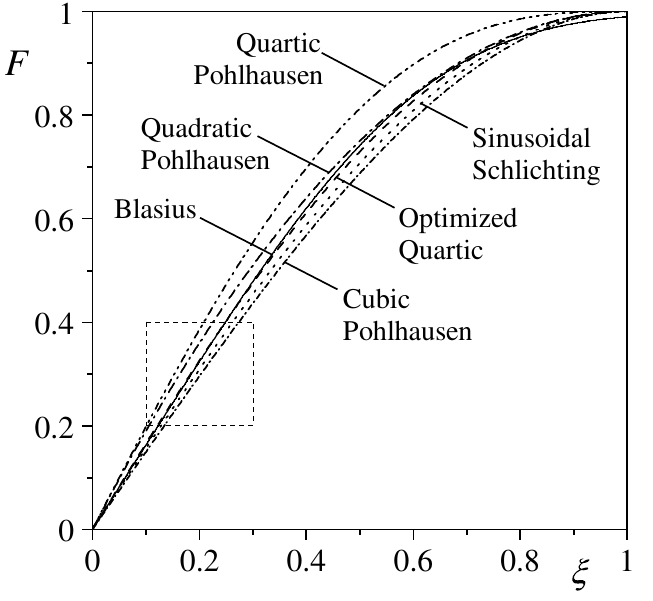}\label{Fig_comparison_of_piecewise_profiles_a}}
	\subfloat[]{\includegraphics[width=0.49\textwidth]{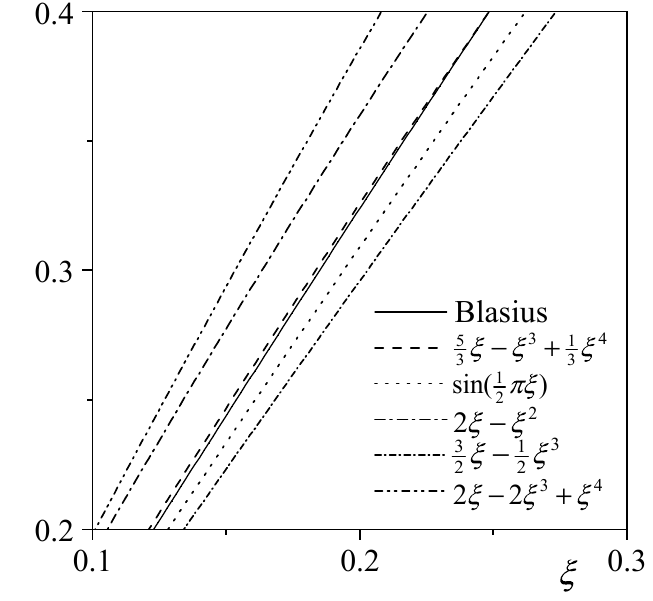}\label{Fig_comparison_of_piecewise_profiles_b}}
	\caption{ Comparison of five analytic approximations for the Blasius solution (solid line) including Pohlhausen's quadratic, cubic, and quartic polynomials (chained lines) as well as Schlichting's sinusoidal (dotted) and the optimized quartic (dashed) profiles. These are shown across (\textit{a}) the boundary layer and (\textit{b}) a designated quadrant where individual deviations from the Blasius curve are magnified.}
	\label{Fig_comparison_of_piecewise_profiles}
\end{figure}

As it turns out, although the fifth condition holds true as $y\to \infty ,$ it proves to be unsuitable at the edge of the boundary layer, where the velocity gradient continues to change.  In fact, a sufficiently resolved numerical solution of the Blasius equation with respect to the boundary-layer coordinate $\xi$ returns a curvature of $-0.7085$, which is not yet zero, but rather of order unity (\cref{Fig_comparison_of_piecewise_profiles}).  
Subjecting the quartic polynomial to the fifth constraint, which happens to be off by one order of magnitude, partly explains its tendency to overpredict several boundary-layer properties relative to its lower-order models (cf. \tblref{Table_coefficient_comparison}).  It also leads to a higher initial slope at $y=\xi =0$, as one may readily infer from \cref{Fig_comparison_of_piecewise_profiles_a}; therein, the piecewise analytic profiles are superimposed on the Blasius line. In contrast, Schlichting's sinusoidal and the optimized quartic profiles display slopes that are closer to the exact Blasius value of 1.630 at $y=0.$ This may be attributed to their wall gradients of $F'(0)= \pi/2 \approx 1.571$ and $5/3\approx 1.667$, which slightly undershoot and overshoot the true slope, respectively.  Other deviations from the Blasius solution can be detected in the two portions of the graph, especially in the magnified inset of \cref{Fig_comparison_of_piecewise_profiles_b}.  Interestingly, the assumption that $u=U$ (and therefore $F=1$ instead of $0.99$) at $y=\delta $ (far right), can be seen to affect all of the piecewise approximations in \cref{Fig_comparison_of_piecewise_profiles_a}, except for the Blasius curve. The latter returns a value of 0.99 at $y=\delta $, which is consistent with the 99\% disturbance basis. This discrepancy, and how to overcome it, will be explored in \cref{Confluence of differential and integral predictions}.

%
\begin{table}

	\begin{center}
		\begin{tabular}{l|c|c|c|c|c|c}
			\toprule
			$F(\xi)=u/U$         & $ F(0) $ &  $ F'(0) $ & $ F''(0) $ & $ F(1) $ & $ F'(1) $ & $ F''(1) $ \\
			\midrule
			$ 2\xi-\xi^2 $  & 0        & 2.000        &  $-2\;\;$      &    1.000     &     0   &     $-2.000$      \\
			$ \tfrac{3}{2}\xi-\tfrac{1}{2}\xi^3 $ 
			& 0        &  1.500        &0       &    1.000     &     0   &     $-3.000$      \\
			$ 2\xi-2\xi^3+\xi^4 $ 
			& 0		   & 2.000        &0        &    1.000    &     0   &      $\;\; 0.000$	\\
			$ \sin(\tfrac{1}{2}\pi\xi) $
			& 0		   & 1.571        &0        &    1.000     &     0   &      $-2.467$	\\
			$ \tfrac{5}{3}\xi-\xi^3+\tfrac{1}{3}\xi^4  $
			& 0		   & 1.667        &0        &    1.000     &     0   &      $-2.000$	\\
			\midrule
			Blasius (1908)
			& 0		   & 1.630       &0        &    0.990 &$0.0904$&      $-0.709$	\\
			\bottomrule
	
	\end{tabular}
    \caption{Endpoint properties of the piecewise analytic velocity profiles and their corresponding Blasius values. All derivatives, including those of the Blasius solution, are taken with respect to $\xi $ instead of $\eta =y\sqrt{U/(2\nu x)} $ to avoid confusion with the Blasius similarity variable. As such, the Blasius edge value, where $u=0.99U$, corresponds to $\eta _{\delta } =\delta \sqrt{U/(2\nu x)} \approx {\rm 3.47188688}.$ }
    \label{Table_endpoint_properties}	
	\end{center}
	\end{table}

For further clarity, the properties and conditions observed by the five piecewise profiles and their derivatives at the endpoints of the viscous domain are summarized in \tblref{Table_endpoint_properties}.  Although not imposed, the negative curvature of the quadratic Pohlhausen profile at $\xi =1$, which matches the curvature of the optimized quartic polynomial of $-2$, appears to be more in-line with the computed Blasius edge curvature than its quartic companion.  This characteristic feature helps to explain, in part, its improved accuracy relative to its quartic form, despite its inability to secure the momentum balance requirement at the wall, i.e., Pohlhausen's fourth condition \cref{Eq_edge_shear_BC}.  In fact, it is only after recognizing these various limitations that one can manage to derive a rationally-optimized quartic polynomial that satisfies Pohlhausen's four essential boundary conditions while judiciously avoiding the fifth requirement.  To make further headway, we posit that the main defect in the classic quartic solution is fundamentally triggered by a prematurely imposed zero shear-stress condition at $y=\delta $.

\subsection{Procedure to develop increasingly accurate polynomials}
\label{Procedure to develop increasingly accurate polynomials}

\noindent The long-standing paradox associated with Pohlhausen's polynomials, which are incorporated into the KP approach, is that their accuracy tends to deteriorate at increasing orders. This behaviour is manifestly established in \tblref{Table_coefficient_comparison}, where their consistently increasing $L_2$ errors with successive increases in their polynomial orders are documented.  This is also evidenced by their increasing spatial deviations from the Blasius solution in \figref{Fig_comparison_of_piecewise_profiles}, including its magnified inset. Although counter-intuitive and thought-provoking paradoxes are somewhat popular in fluid analysis, a formal procedure that overcomes this perplexing behaviour is desirable.

To begin, we opt for a non-dimensional velocity profile that can be approximated by an $ N{\rm th} $-order polynomial of the form
\begin{equation}\label{Eq_generic_series_expansion}
F(\xi)=\sum_{n=1}^{N}{a_n\xi^n}
\end{equation}
\noindent Based on \cref{Eq_nond_displacement_thicknesses}, the non-dimensional displacement and momentum thicknesses can be written as
\begin{equation}\label{Eq_series_expansions}
\begin{split}
\alpha&=\int_{0}^{1}{(1-F)\dd \xi}=1-\sum_{n=1}^{N}\frac{a_n}{n+1}\\
\beta& =\int_{0}^{1}{F(1-F)\dd \xi}=\sum_{n=1}^{N}{\frac{1}{n+1}\left(a_n-\sum_{k=1}^{n-1}{a_ka_{n-k}} \right) } \\
\end{split}
\end{equation}
Polynomials that conform to this series expansion and that satisfy some (or all) of the conditions given by \cref{Eq_noslip,Eq_smoothedge_BC,Eq_wall_momentum_balance_BC,Eq_edge_shear_BC} include, but are not limited to,
\begin{equation}\label{Eq_different_polynomials}
F(\xi)=\xi, \quad F(\xi)=2\xi-\xi^2, \quad F(\xi)=\tfrac{3}{2}\xi - \tfrac{1}{2}\xi^3, 
\quad F(\xi)=2\xi - 2\xi^3+\xi^4
\end{equation}
On the one hand, when Pohlhausen's physical requirements \cref{Eq_noslip,Eq_smoothedge_BC,Eq_wall_momentum_balance_BC,Eq_edge_shear_BC} are applied on $F(\xi)$, they translate into
\begin{equation}\label{Eq_BC_Pohlhausen}
F(0)=F^{\prime\prime}(0)=0, \quad F(1) =1,\quad F'(1)=F^{\prime \prime}(1)=0 
\end{equation}
On the other hand, a robust numerical solution of the Blasius equation predicts, as shown in \tblref{Table_endpoint_properties}, 
\begin{equation}\label{Eq_BC_Blasius}
\Fbar(0)=\Fbar^{\prime\prime}(0)=0, \quad \Fbar(1) =0.99,\quad \Fbar'(1)\approx0.090357643, \quad \Fbar^{\prime \prime}(1)\approx-0.70853762 
\end{equation}
It can thus be seen that the $ F^{\prime \prime}(1)=0 $ constraint, which is secured by Pohlhausen's quartic solution, entails a non-negligible error. To overcome this limitation, we hypothesize that the $ F^{\prime \prime}(1)=0$ requirement must be released. The degree of freedom thereby gained can be leveraged while taking into account the direct relations identified in \cref{Direct connection to an assumed velocity profile}, i.e., to produce the closest values of $\{\alpha, \beta, s\}$ to the problem's exact solution.

To illustrate this procedure, we choose a quartic polynomial that satisfies all conditions in \cref{Eq_BC_Pohlhausen} except $ F^{\prime \prime}(1)=0 $.  Despite the penalty that the $F'(1)=0$ assumption engenders, it is retained here in observance of the Pohlhausen tradition of suppressing the gradient of the velocity at the edge of the domain.  Such a profile consists of 
\begin{equation}\label{Eq_MX_conforming_profile}
F(\xi)=s\xi+(4-3s)\xi^3+(2s-3)\xi^4
\end{equation}
Then using the reduced relations specified in \cref{Eq_nond_displacement_thicknesses}, the non-dimensional disturbance and momentum thicknesses can be expressed explicitly as functions of $s$. We get
\begin{equation}\label{Eq_alpha_beta_functions_of_s}
\alpha=\frac{3}{5}-\frac{3}{20}s\quad \quad \text{and} \quad \quad
\beta=\frac{4}{35}+\frac{13}{210}s-\frac{19}{630}s^2 
\end{equation}
At the time the KP approach was being developed \citep{Toepfer1912}, the classical Blasius solution predicted $ \bar{\alpha}\approx0.344$, $\bar{\beta}\approx0.133$ and $ \sbar\approx1.660$ \citep{Schlichting1979,Fox2004}. To minimize the overall error relative to the traditionally reported Blasius values, it is useful to define the following objective function:
\begin{equation}\label{Eq_target_function}
\mathcal{R}=\left(\frac{\alpha-{\bar\alpha}}{{\bar\alpha}}\right)^2 + \left(\frac{\beta-{\bar\beta}}{{\bar\beta}}\right)^2
+\left(\frac{s-\sbar}{\sbar}\right)^2
\end{equation}
We have chosen to constrain the total relative residual of $\alpha$, $\beta$ and $s$, because these parameters have been shown in \cref{Direct connection to an assumed velocity profile} to be the most influential in prescribing the remaining boundary-layer properties. At this juncture, we may substitute \cref{Eq_alpha_beta_functions_of_s} into \cref{Eq_target_function} and express $\mathcal{R}$ solely in terms of $s$.  This enables us to minimize $\mathcal{R}(s)$ analytically, namely, by taking,
\begin{equation}\label{Eq_minimization_of_target_function}
\left. \frac{\dd \mathcal{R}}{\dd s} \right|_{s=s\raisebox{+.05ex}{\footnotesize{*}}}=0\quad \text{where} \quad \mathcal{R}(s)=0.0514s^4-0.211s^3+0.833s^2-1.985s+1.574
\end{equation}
Thus, by differentiating the objective function with respect to $s$, one is rationally guided to select $s\raisebox{+.15ex}{{*}} \approx 1.6794\approx 5/3$, which is still $2.2\%$ higher than the Blasius slope of $\sbar\approx 1.6304$ and, unsurprisingly, lower than the classic Pohlhausen slope of $2$ by precisely $20\%$. Evidently, the use of $s\raisebox{+.15ex}{{*}}$ leads to an excellent overall approximation, albeit slightly different from the Blasius constant. 
Lastly, by inserting this trajectory back into \cref{Eq_MX_conforming_profile} and simplifying, we arrive at
\begin{equation}\label{Eq_amended_MX_profile}
F(\xi)=\frac{5}{3}\xi-\xi^3+\frac{1}{3}\xi^4 
\end{equation}
It is a simple matter to verify that this rationally-optimized profile satisfies Pohlhausen's four fundamental boundary conditions, while dodging the unrealistic $ F^{\prime \prime}(1)=0 $ constraint.  One may also confirm that this solution leads to reliable estimates for the boundary-layer thickness and skin friction, namely, $(\delta/x)\sqrt {R{e_x}} \approx4.993 $ and $ C_f\sqrt {R{e_x}}\approx0.668 $.  These estimates differ from the classical Blasius values of $5.0$ and $0.664$ by $0.13$ and $0.53$\%, respectively.  Compared to other profiles featured in \tblref{Table_coefficient_comparison}, its $L_2$ error of $0.008$, evaluated over its finite interval, is another favourable metric: it reflects an order of magnitude reduction relative to its predecessors.
As an additional side benefit, the simplicity of \cref{Eq_amended_MX_profile} enables us to extract closed-form expressions for the normal velocity component and its peak value. As outlined in \cref{Eq_v_velocity_derivation3,Eq_v_velocity_max}, continuity may be used to retrieve 

\begin{equation}
\label{Eq_v_velocity_MX_value}
\frac{v}{U} =\left(\xi ^{2} -\frac{3}{2} \xi ^{4} +\frac{4}{5} \xi ^{5} \right)\frac{(a/2)}{\sqrt{Re_{x} } } \approx \left(\xi ^{2} -\frac{3}{2} \xi ^{4} +\frac{4}{5} \xi ^{5} \right)\frac{2.9178}{\sqrt{Re_{x} }} 
\end{equation} 
As usual, the maximum value of $v$ occurs at the edge of the layer, namely,
\begin{equation}
\label{Eq_v_velocity_MX_max_value}
\left. \frac{v}{U} \right|_{\xi=1 } =\frac{(7a/40)}{\sqrt{Re_{x} } } \approx \frac{0.8738}{\sqrt{Re_{x} } } 
\end{equation} 
In relation to the general form prescribed by \cref{Eq_v_velocity_max,Eq_v_velocity_solution}, we recover here
\begin{equation}
\label{Eq_v_velocity_G_value}
G(\xi) =a \left( \frac{\xi ^{2}}{2} -\frac{3}{4} \xi ^{4} +\frac{2}{5} \xi ^{5} \right) \quad \text{and} \quad G(1)= \frac{7a}{40}\approx 0.8738
\end{equation} 
Despite its simplicity, the ability of \cref{Eq_v_velocity_MX_value} to predict its Blasius analogue is quite satisfactory. Recalling from \cref{Eq_v_max_coefficient} that the normal velocity corresponding to the classic Blasius solution is proportional to $\alpha a/2$, we get
\begin{equation}
\label{Eq_v_velocity_Blasius_value}
\left. \left(\frac{v}{U} \right)_{} \right|_{\xi=1 } \approx \frac{0.86}{\sqrt{Re_{x} } } \quad\quad \text{ (classic Blasius value)}
\end{equation} 
It is reassuring that the prediction based on the optimized quartic profile is off by less than 1.6\%.  

\subsection{Extension to higher-order polynomial approximations} \label{Extension to higher-order polynomial approximations}
\noindent Pursuant to this analysis, higher-order polynomials can be constructed by securing additional boundary conditions and properties associated with the Blasius problem. For example, quintic, sextic, and septic polynomials can be judiciously constructed to mirror the behaviour of the Blasius solution at the two endpoints of the viscous layer.  These can be arrived at using a similar optimization technique to the one pursued to obtain a quartic profile, namely, by solving for the optimal values of $s\raisebox{+.15ex}{{*}}$ and other coefficients that stand to minimize the total residual of several additional constraints. 
To sketch this process, we start with $N=5$ in \cref{Eq_generic_series_expansion} and then select three foundational conditions that must be fixed in the optimization procedure in order to properly capture the character of the Blasius solution. We find it essential to set:
\begin{equation}\label{Eq_Three_BCs_for_polynomials}  
\begin{gathered}
F(0)=F''(0)=0 \quad\text{and} \quad F(1)=0.99\\
\end{gathered}
\end{equation} 
These immutable equalities enable us to determine three of the unknown constants, $a_0=a_2=0$, and $a_3=0.99-a_1-a_4-a_5$.  As for the remaining parameters, $a_1$, $a_4$, and $a_5$, they can be evaluated in such a way to minimize the objective function $\mathcal{R}$ defined in \cref{Eq_target_function}. Thus, by simultaneously solving the system of constraints,
\begin{equation}\label{Eq_general_minimization_of_target_function}
\frac{\dd \mathcal{R}}{\dd a_1}=0\quad \frac{\dd \mathcal{R}}{\dd a_4} =0 \quad \text{and} \quad \frac{\dd \mathcal{R}}{\dd a_5}=0
\end{equation}
we extract $a_1=s\raisebox{+.15ex}{{*}}\approx1.630398$, $a_4\approx-2.041276$ and $a_5\approx1.291600$. Our optimized quintic polynomial becomes
\begin{equation}\label{Eq_quintic_poly}
F(\xi)\approx 1.6304\xi+0.1093\xi^3 -2.0413\xi^4+1.2916\xi^5
\end{equation}

\noindent Note that the value of $s\raisebox{+.15ex}{{*}}$ here has essentially converged onto the exact Blasius trajectory. Moreover, despite its simple four-term composition, \cref{Eq_quintic_poly} proves to be quite accurate, as it exhibits an $L_2$ error of $0.002$. In addition to providing a simple expression that can be conveniently integrated and differentiated, it reproduces the fundamental boundary-layer characteristics, precise in at least four decimal places, with a maximum relative error of $0.0012\%$ in the estimation of $b$. Its incorporation within the KP approach leads to the following refined values, accurate in all digits displayed:
\begin{equation}\label{Eq_predictions_using_quintic_poly}  
\begin{gathered}
a=4.9100 \quad b=0.6641 \quad c=1.7208 \quad \alpha=0.3505 \quad \beta=0.1353 \quad \text{and}\quad H=2.5911 \\
\end{gathered}
\end{equation}

Extending this procedure further to a sextic polynomial can be achieved with relative ease.  Using the same fixed conditions as before, the additional degree of freedom that we gain with $N=6$ can be invested in minimizing an additional constraint such as $F'(1)=\Fbar'\approx 0.090357643$ in \cref{Eq_BC_Blasius}.  Our objective function becomes: 
\begin{equation}\label{Eq_target_function_for_sextic_poly}
\mathcal{R}=\left(\frac{\alpha-{\bar\alpha}}{{\bar\alpha}}\right)^2 + \left(\frac{\beta-{\bar\beta}}{{\bar\beta}}\right)^2
+\left(\frac{s-\sbar}{\sbar}\right)^2+\left[ \frac{F'(1)-\Fbar'}{\Fbar'} \right]^2
\end{equation}
As before, we first extract $a_0=a_2=0$ and $a_3=0.99-a_1-a_4-a_5-a_6$.  The remaining four constants $\{a_1, a_4,a_5,a_6\}$ can be optimally determined such that
\begin{equation}\label{Eq_general_minimization_of_target_function}
\frac{\dd \mathcal{R}}{\dd a_1}=\frac{\dd \mathcal{R}}{\dd a_4} =\frac{\dd \mathcal{R}}{\dd a_5} =\frac{\dd \mathcal{R}}{\dd a_6}=0
\end{equation}
This system can be solved straightforwardly. One gets
\begin{equation}\label{Eq_sextic_poly}
F(\xi)\approx 1.6304\xi+0.4950\xi^3-3.8897\xi^4+3.9920\xi^5-1.2377\xi^6
\end{equation}
With an $L_2$ error of $0.001758$, the minor improvement that accompanies the sextic approximation is reflected in a maximum error of $0.0019\%$, which is accrued in the estimation of $b$; all characteristic properties remain, within the number of digits displayed, no different from those reported in \cref{Eq_predictions_using_quintic_poly}. Hence, the use of an optimized sixth-order versus a fifth-order polynomial does not seem to be warranted, as the overall benefit remains minimal, except in the ability to better represent the endpoint conditions.   

A septic polynomial, however,  has the advantage of satisfying eight conditions identically, and these can absorb the values of the Blasius velocity function and its first three derivatives at both ends of the domain. In other words, a septic polynomial can be constructed either through optimization, or in a manner to reproduce $F'(0)=\sbar$ precisely, along with seven other Blasius constraints.  After some effort, we find that an assortment of eight physical requirements leads to a reasonably accurate solution.  
For example, using $N=7$ in \cref{Eq_generic_series_expansion}, the following eight conditions can be imposed:
\begin{equation}\label{Eq_Eight_BCs_for_polynomials}  
\begin{gathered}
F(0)=0 \quad F'(0)=\sbar \quad  F''(0)=0 \quad F'''(0)=0\quad F(1)=0.99\\
F'(1)\approx 0.0903576430617476 \quad F''(1)\approx -0.7085376161268071 \\
F'''(1)\approx 4.4777040873935454\\
\end{gathered}
\end{equation} 

\noindent These lead to $a_0=a_2=a_3=0$, $a_1=\sbar$, with the remaining constants yielding
\begin{equation}\label{Eq_coefficients_for_polynomials}  
\begin{gathered}
a_4=0.99-\sbar-a_5-a_6-a_7\quad a_5=3\sbar-2a_6 -3a_7-3.86964 \\
a_6=9.1843-6\sbar-3a_7 \quad\text{and}\quad a_7=10\sbar-16.7331 \\
\end{gathered}
\end{equation} 
As such, backward substitution into \cref{Eq_generic_series_expansion} returns
\begin{equation}\label{Eq_septic_polynomial_reduced}  
\begin{gathered}
F(\xi)\approx \sbar \xi-1.83095 \xi^4 + 0.93048 \xi^5+0.68916 \xi^6-0.42908 \xi^7\\
\end{gathered}
\end{equation}

\noindent Despite its simple five-term composition, \cref{Eq_septic_polynomial_reduced} proves to be quite accurate, with an $L_2$ error of $0.002$, which is comparable to the optimized quintic and sextic solutions. Nonetheless, although it matches the Blasius solution at the endpoints precisely, it reproduces the boundary-layer characteristics with a maximum error of 0.36\%. For example, it predicts the following parameters and their corresponding relative errors (parenthetically): 
\begin{equation}\label{Eq_predictions_using_septic_poly}  
\begin{gathered}
a=4.9143\; (0.09\%) \quad b=d=0.6635 \;(0.09\%) \quad c=1.7254 \; (0.27\%) \\
\alpha=0.3511 \; (0.18\%)  \quad \beta=0.1350 \; (0.18\%)   \quad\text{and}\quad  H=2.600 \; (0.36\%) \\
\end{gathered}
\end{equation}

Even at this high order, the benefit of optimization is clear. Instead of imposing the endpoint constraints on the third derivatives in \cref{Eq_Eight_BCs_for_polynomials}, one may exclude $ F'''(0)=0$ and $F'''(1)\approx4.4777$ in favour of converting an additional Blasius condition into an objective constraint in \cref{Eq_target_function_for_septic_poly}. The latter may be augmented by the $F''(1)=\Fbar''\approx-0.70853762$ target requirement, namely,

\begin{equation}\label{Eq_target_function_for_septic_poly}
\mathcal{R}=\left(\frac{\alpha-{\bar\alpha}}{{\bar\alpha}}\right)^2 + \left(\frac{\beta-{\bar\beta}}{{\bar\beta}}\right)^2
+\left(\frac{s-\sbar}{\sbar}\right)^2+\left[\frac{F'(1)-\Fbar'}{\Fbar'}\right]^2+\left[ \frac{F''(1)-\Fbar''}{\Fbar''} \right]^2
\end{equation}
As before, a systematic solution leads to $a_0=a_2=0$ and $a_3=0.99-a_1-a_4-a_5-a_6-a_7$. This is followed by minimization with respect to $\{a_1, a_4,a_5,a_6,a_7\}$, which enables us to retrieve
\begin{equation}\label{Eq_septic_poly_optimized}
F(\xi)\approx 1.6304\xi+0.6840\xi^3-5.0756\xi^4+6.5605\xi^5-3.5729\xi^6+0.7636\xi^7
\end{equation}
In this case, the extremum corresponds to a zero residual for which $a_1=\sbar$ identically.  Surprisingly, however, the $L_2$ error of $0.0019$ does not decrease relative to the sextic approximation. Nonetheless, by exhibiting an overall error that is comparable to that of the quintic solution, the septic expression is capable of restoring the boundary-layer properties with a maximum $0.0012\%$ discrepancy in $\beta$. And though its overall error slightly exceeds that of the sextic solution, its main advantage stands in its ability to satisfy the endpoint values more accurately. It also captures the boundary-layer properties more precisely than its five-term septic analogue, \cref{Eq_septic_polynomial_reduced}, mainly because the latter is prescribed by the entire set of conditions in \cref{Eq_Eight_BCs_for_polynomials}, notwithstanding the values of $\alpha$ and $\beta$. The inclusion of $\alpha$ and $\beta$ as target functions in an optimization procedure is therefore essential.

\subsection{Improved effectiveness in thermal analysis}\label{Improved effectiveness in thermal analysis}
\begin{figure}
	\centering
	\includegraphics[width=0.7\textwidth]{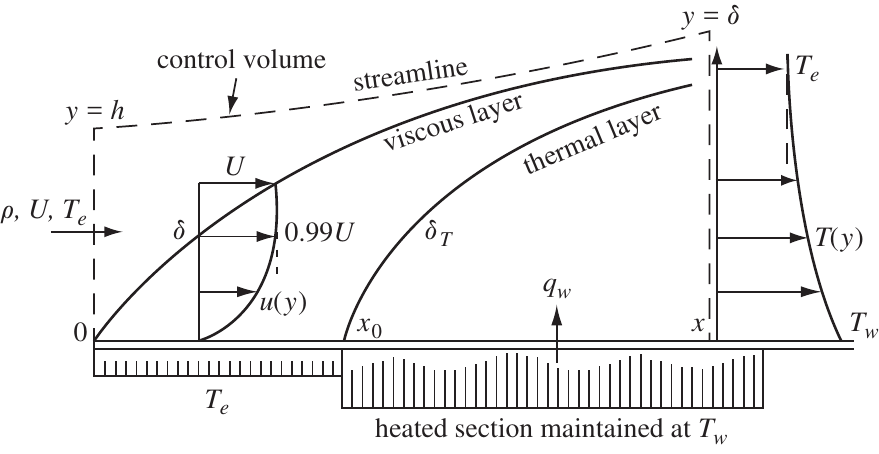}	
	\caption{ Sketch of the two-pronged viscous and thermal analysis based on the KP integral approach for flow over a flat plate. A heated wall is shown for which $\delta_{T} \leq \delta $. }
	\label{Fig_CV_thermalBL_analysis}
\end{figure}

\noindent Having illustrated the utility of a refined formulation in a viscous-flow application, we can now proceed by applying the KP integral approach in conjunction with an energy balance to retrieve valuable estimates of the problem's heat transfer and thermal boundary-layer characteristics. A sketch of our domain is provided in \figref{Fig_CV_thermalBL_analysis}, where a section of the plate is heated, for $x\ge x_{0} $, to a constant temperature $T_{w} $. As usual, the edge temperature $T_{e} $ corresponds to the freestream conditions. In this setting, a thermal boundary layer, $\delta _{T} $, will begin to develop starting at $x_{0}$. As shown by \citet{White2006}, the flat-plate configuration can be modelled using 
\begin{equation} \label{Eq_energy_balance} 
q_{w} \approx \frac{\dd}{\dd x} \left[\int _{0}^{\delta _{T} }\rho c_{p} u\left(T-T_{e} \right)\dd y \right] =-k\left. \frac{\partial T}{\partial y} \right|_{y=0}
\end{equation} 
where $ q_{w}, c_p,$ and $k$ are used to designate the wall heat flux, specific heat, and thermal conductivity, respectively.

\begin{figure}
	\subfloat[]{\includegraphics[width=0.49\textwidth]{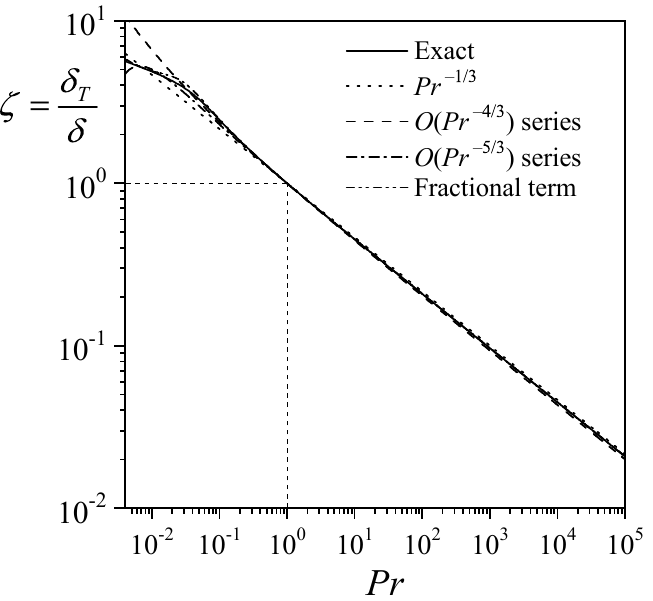}\label{Fig_ThermalBL-Predictions_a}}
	\subfloat[]{\includegraphics[width=0.49\textwidth]{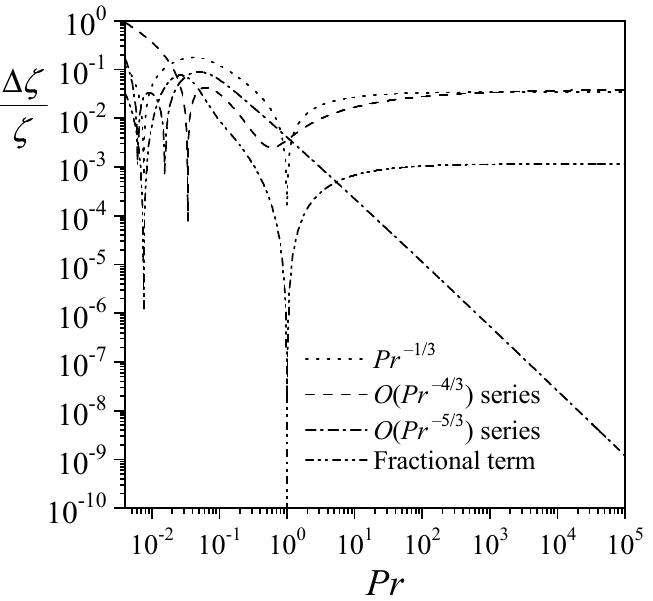}\label{Fig_ThermalBL-Predictions_b}}
	\caption{ Comparison of several analytic approximations for $\zeta =\delta _{T} /\delta $ over a wide range of Prandtl numbers with (\textit{a}) clear convergence characteristics for all models on the $Pr^{-1/3} $ expression for $Pr>0.3$ and (\textit{b}) well-behaved relative errors, particularly with the optimized quartic polynomial profile.}
	\label{Fig_ThermalBL-Predictions}
\end{figure}

To make further headway, an assumed velocity profile must be selected, say from \cref{Eq_Pohlhausen,Eq_Schlichting,Eq_Majdalani-Xuan}, and then paired with a shape-preserving temperature approximation.  To draw a meaningful comparison, two cases will be considered: the first will apply Pohlhausen's quadratic polynomial (with the smallest $L_2$ error in \tblref{Table_coefficient_comparison}), and the second will repeat the analysis using the optimized quartic polynomial in \cref{Eq_amended_MX_profile}.  In the interest of brevity, the thermal boundary layer will be permitted to start at the leading edge, with $\delta =\delta _{T} =0$ at $x=x_0=0$ in \cref{Fig_CV_thermalBL_analysis}. Using a similar quadratic polynomial to describe the temperature distribution within the thermal layer \citep{White2006}, we can write  
\begin{equation} \label{Eq_quadratic_temp} 
\frac{T-T_{e}}{T_{w} -T_{e}} \approx 1-\left(\frac{2y}{\delta _{T} } -\frac{y^{2} }{\delta _{T}^{2} } \right)
\end{equation} 
When this approximation is substituted into \eqref{Eq_energy_balance}, integrated over $\delta _{T} $, and rearranged, one is left with 
\begin{equation} \label{Eq_quadratic_energy_balance} 
\frac{q_{w} }{T_{w} -T_{e} } \approx \frac{\dd}{\dd x} \left[\rho c_{p} U\delta \left(\frac{\zeta ^{2} }{6} -\frac{\zeta ^{3} }{30} \right)\right]= \frac{2k}{\zeta \delta }  
\end{equation} 
where $\zeta =\delta _{T} /\delta $ captures the thermal-to-viscous ratio of boundary-layer thicknesses. At this juncture, we may substitute $\delta (x)=ax^{1/2} /\sqrt{U/\nu } $ from \cref{Eq_a_constant}, with $a=\sqrt{30} \approx 5.477$ from \tblref{Table_coefficient_comparison}, to obtain
\begin{equation} \label{Eq_zeta_quartic1} 
\zeta ^{3} -\frac{\zeta ^{4} }{5} =\frac{24\alpha_T }{a^{2} \nu } =\frac{0.8}{Pr} 
\end{equation} 
where $\alpha_T=k/(\rho c_p)$ is the thermal diffusivity.  An asymptotic solution to this polynomial, valid over a practical range of Prandtl numbers, may be readily extracted. One finds
\begin{equation} \label{Eq_zeta_quartic2} 
\zeta =0.9283Pr^{-1/3} +0.05745Pr^{-2/3} +0.01067Pr^{-1} +{{O}}(Pr^{-4/3} )\approx Pr^{-1/3}
\end{equation} 
As shown with a dashed line in \cref{Fig_ThermalBL-Predictions}, our ${O}(Pr^{-4/3} )$ series expansion agrees well with the numerical solution (solid line) as well as the simplest, one-term, $Pr^{-1/3} $ expression (dotted line). With $\zeta $ and $\delta $ in hand, we can proceed to construct the local Nusselt number using \eqref{Eq_quadratic_energy_balance}. We get
\begin{equation} \label{Eq_Nusselt_no_quadratic} 
Nu_{x} =\frac{q_{w} x}{k(T_{w} -T_{e} )} =\frac{2k(T_{w} -T_{e} )x}{\zeta \delta k(T_{w} -T_{e} )} =\frac{2x}{\zeta \delta } \approx 0.365Re_{x}^{1/2} Pr^{1/3}  
\end{equation} 
In this case, the estimated $Nu_x$ coefficient overshoots the correct value for constant wall temperature analysis of a flat plate by 10\% (the correct constant being 0.332).  

The classical KP procedure just described can be repeated using \cref{Eq_Majdalani-Xuan}. First, \eqref{Eq_quadratic_temp} may be replaced by the optimized quartic polynomial estimate for the temperature such as
\begin{equation} \label{Eq_quartic_temp} 
\frac{T-T_{e}}{T_{w} -T_{e}} \approx 1-\left(\frac{5y}{3\delta _{T} } -\frac{y^{3} }{\delta _{T}^{3} } +\frac{y^{4} }{3\delta _{T}^{4} } \right)
\end{equation} 
Inserting this expression into the energy balance integral \eqref{Eq_energy_balance}, we arrive at
\begin{equation} \label{Eq_quartic_energy_balance1} 
\frac{q_{w}}{\rho c_{p} \left(T_{w} -T_{e} \right)} \approx \frac{\dd}{\dd x} \left[\int _{0}^{\delta _{T} }U\left(\frac{5y}{3\delta } -\frac{y^{3} }{\delta _{}^{3} } +\frac{y^{4} }{3\delta _{}^{4} } \right)\left[1-\left(\frac{5y}{3\delta \zeta } -\frac{y^{3} }{\zeta ^{3} \delta _{}^{3} } +\frac{y^{4} }{3\zeta ^{4} \delta _{}^{4} } \right)\right]\dd y \right]=\frac{5\alpha_T}{3\delta \zeta } 
\end{equation} 
Further expansions, integrations, and simplifications leave us with
\begin{equation} \label{Eq_quartic_zeta_equation} 
	\dfrac{3\delta \zeta }{5} \dfrac{\dd}{\dd x} \left[\left(\dfrac{4\zeta ^{2} }{27} -\dfrac{\zeta ^{4} }{56} +\dfrac{11\zeta ^{5} }{3240} \right)U\delta \right]=\alpha_T\quad     \text{or}    \quad\zeta ^{3} -\dfrac{27\zeta ^{5} }{224} +\dfrac{11\zeta ^{6} }{480} =\dfrac{0.90238}{Pr} 
\end{equation}
\noindent The resulting expression can be put in the form $\zeta ^{3} -\alpha _{0} \zeta ^{5} +\beta _{0} \zeta ^{6} =\gamma _{0}^{3} Pr^{-1} $ with $\alpha _{0} ={\tfrac{27}{224}},$ $\beta _{0} ={\tfrac{11}{480}},$ and $\gamma _{0} =\sqrt[{3}]{{\tfrac{379}{420}} } $.  Here too, a robust approximation can be found, namely, $\zeta =\gamma _{0} Pr^{-1/3} +{\tfrac{1}{3}} \alpha _{0} \gamma _{0}^{3} Pr^{-1} -{\tfrac{1}{3}} \beta _{0} \gamma _{0}^{4} Pr^{-4/3} +{O}(Pr^{-5/3} ).$ More visibly, we get
\begin{equation} \label{Eq_quartic_zeta_soln} 
\zeta =0.96634Pr^{-1/3} +0.0362564Pr^{-1} +0.00666116Pr^{-4/3} +{O}(Pr^{-5/3} )\approx Pr^{-1/3}
\end{equation}
As illustrated with a chained (dash-dot) line in \figref{Fig_ThermalBL-Predictions}, this three-term perturbation series of ${O}(Pr^{-5/3} )$ exhibits a steadily diminishing error with successive increases in the Prandtl number.  It can be relied upon so long as $Pr>0.0045$, knowing that its error drops precipitously below 1\% for $Pr>0.45$. Another `fractional' approximation that outperforms the ${O}(Pr^{-5/3} )$ expansion in the $0.004<Pr<5.35962$ range can also be achieved using 
\begin{equation} \label{Eq_quartic_zeta_alternative_soln} 
\zeta =\left(\frac{1620Pr^{1/3} -9752Pr-385}{675Pr^{1/3} -3360Pr-154} \right)\frac{1}{3Pr^{1/3} } 
\end{equation}
These solutions and their relative errors are showcased in \figref{Fig_ThermalBL-Predictions} along with the one-term theoretical $Pr^{-1/3} $.  Despite the transcendental character of these equations, it is reassuring that they all return a root of $\zeta =1$ for $Pr=1$, in compliance with the Reynolds analogy. Moreover, their differences become graphically imperceptible for $Pr>0.3$ in \figref{Fig_ThermalBL-Predictions_a} as their relative errors in \figref{Fig_ThermalBL-Predictions_b} drop to 4\% or lower.  These observations confirm the viability of using $\zeta \approx Pr^{-1/3} $, which entails a less than 3.3\% error, for all gases and fluids with $Pr>0.7$.

To complete our thermal analysis in the wake of $\zeta $ and $\delta $, and recalling that $a= \delta \sqrt{Re_{x} } /x=\sqrt{{\tfrac{9450}{379}} } \approx 4.9934$ from \tblref{Table_coefficient_comparison}, the local Nusselt number can be deduced systematically by taking
\begin{equation} \label{Eq_Nusselt_no_quartic1} 
	Nu_{x} =\dfrac{q_{w} x}{k(T_{w} -T_{e} )} =k\left(T_{w} -T_{e} \right)\dfrac{5}{3\delta \zeta } \dfrac{x}{k(T_{w} -T_{e} )} =\dfrac{5x}{3\delta \zeta }     
\end{equation}
and so    
\begin{equation} \label{Eq_Nusselt_no_quartic2} 
	\ Nu_{x} =\dfrac{5Re_{x}^{1/2} Pr^{1/3} }{3a} =\dfrac{5Re_{x}^{1/2} Pr^{1/3} }{3\sqrt{{\tfrac{9450}{379}} } } \approx 0.3338Re_{x}^{1/2} Pr^{1/3} 
\end{equation}
\noindent The resulting coefficient is only 0.5\% higher than the `exact' coefficient of 0.332.  As such, the use of an optimally-constructed quartic profile in lieu of Pohlhausen's is greatly beneficial. Clearly, simple integral techniques that are based on rationally-conceived velocity and temperature profiles can yield valuable approximations for either local or global friction and heat transfer coefficients, with degrees of accuracy that range from less than 1\%, as shown in this study, to 10\% or more, depending on the precision of the assumed profiles. 

In leaving this section, we note that although the same procedure can be extended to include pressure gradient and compressibility effects, such developments fall outside the scope of this work. Our interest will shift, instead, to the presentation of more accurate and continuously analytic solutions that remain uniformly valid as $\xi \rightarrow \infty$.
 
\section{Development of compact, uniformly valid solutions for the Blasius problem}\label{Development of a continuous solution for the Blasius problem}

\subsection{Benchmark for flat-plate flow analysis}

\noindent Shortly after the unveiling of the foundational boundary-layer equations by \citet{Prandtl1904}, \citet{Blasius1908} may have produced one of the most transformational similarity solutions in fluid mechanics. Although his conversion begins with the reduced Navier-Stokes equations, the ensuing work has proven to be an invaluable tool for the practical treatment of viscous flows. Not only has it provided deeper insight into boundary-layer effects, it has truly ushered in a series of similarity solutions and an era of unprecedented growth in the field. 

In short, by introducing a similarity variable of the form $\eta =y\sqrt{{U}/{(2\nu x)} }$, Blasius manages to reduce Prandtl's boundary-layer equations to a simple nonlinear differential equation for the flat-plate flow problem outlined in \cref{Fig_CV-flat-plate}.  This equation would later serve as a permanent benchmark for testing new methods of analysis and a lasting beacon in the quest for new similarity solutions. Note that the factor `2' under the radical is adopted here for convenience. Then using the same notation as before, we take $\psi =\sqrt{2\nu Ux} f(\eta )$ to be the parental stream function. The corresponding velocities become
\begin{equation} \label{Eq_Blasius_velocities} 
u=\dfrac{\partial \psi }{\partial y} =U \dot{f}=U \Fbar \;   \quad  \; v=-\dfrac{\partial \psi }{\partial x} =\dfrac{U}{\sqrt{2Re_{x} } } (\eta \dot{f}-f)=\dfrac{U}{\sqrt{2Re_{x} } } \left(  \dfrac{a}{\sqrt{2}}\xi \Fbar -\int _{0}^{\xi} \Fbar \dd \xi \right)
\end{equation} 
where overdots denote differentiation with respect to $\eta $, which is deliberately used to avoid confusion with $\xi$. Except for $\Fbar(\xi )=u/U=\dot{f}$, dotted derivatives with respect to $\eta$ differ from those referred to $\xi=\eta/\eta_{99\%} =\eta\sqrt{2}/a$.  When these transformations are substituted into Prandtl's equations, a widely celebrated third-order differential equation is recovered, namely,
\begin{equation} \label{Eq_Blasius_equation} 
\dddot{f}+f\ddot{f}=0 \quad \text{or} \quad {f'''}+\frac{a}{\sqrt{2}} f {f''}=0
\end{equation} 
As for the boundary conditions, the velocity adherence at the wall, $u(x,0)=v(x,0)=0$, and the freestream-merge condition, $u(x,\infty )=U,$ readily convert through \eqref{Eq_Blasius_velocities} into
\begin{equation} \label{Eq_Blasius_BCs} 
	\dot{f}(0)=f(0)=0\quad \dot{f}(\infty )=1  \quad  \text{or} \quad   {f'}(0)=f(0)=0\quad {f'}(\infty )=1 
\end{equation} 

\subsection{Series expansions for the Blasius problem}
\noindent So far, the Blasius equation has only yielded exact solutions that are cast in the form of infinite series. Examples include the Homotopy Analysis Method (HAM) series by \citet{Liao1997,Liao1999a,Liao1999b} and \citet{Liao2002}, as well as the Adomian decomposition of an exponentially-transformed Blasius variable by \citet{Ebaid2013}.  Although \citet{Blasius1908} provides asymptotic approximations for small and large $\eta $, his series expansion exhibits a finite convergence radius of $0\le \eta <4.0234644935,$ beyond which the solution diverges.  The Blasius expansion also requires knowledge of the initial gradient of the axial velocity with respect to $\eta $, denoted here as $\sigma=\ddot{f}(0)\approx0.46959999,$\footnote{This value shifts to  $\sigma /\sqrt{2} \approx 0.3320573362151963$  when the similarity variable and stream function are written slightly differently as  $\eta =y\sqrt{U/(\nu x)} $  and  $\psi =\sqrt{\nu Ux} f(\eta )$. The corresponding Blasius equation becomes  $\dddot{f}+{\tfrac{1}{2}} f\ddot{f}=0$,  with the same three conditions as in \eqref{Eq_Blasius_BCs}.  Moreover, the initial slope changes to  $\sbar=\Fbar'(0) =\sigma a/\sqrt{2} \approx1.630398038629397$,  when written as the wall derivative of the velocity function  $\Fbar(\xi )=u/U$  of \cref{Direct evaluation of properties using piecewise velocity approximations} (cf. \tblref{Table_coefficient_comparison}), where   $\xi =y/\delta$ .} a value that is often termed `connection parameter' or `Blasius constant.'  

Naturally, in view of its relevance to a wide variety of problems, the asymptotic behaviour of the Blasius solution has been the subject of numerous investigations. Some of these inquiries have devoted themselves to its existence and uniqueness characteristics \citep{Weyl1942,Callegari1968,Callegari1978,Fang2008}, including its tri-pole singularity inside its semi-infinite domain \citep{Punnis1956a,Punnis1956b,Boyd1999}.  Others have attempted to overcome its deceptive singularity using Pad\'{e} approximants \citep{Boyd1997,Ahmad2009,Peker2011} or Crocco variables to relocate the singularity to a more convenient outpost \citep{Callegari1968,Callegari1978,Wang2004}.

In practice, the persistent efforts to solve the Blasius equation have led to vastly dissimilar algebraic expressions.  These extend from simple, piecewise approximations valid up to the edge of the boundary layer, i.e., $0\le \eta \le \eta _{\delta } $, where $\eta _{\delta } =\delta _{99\% } \sqrt{U/(2\nu x)} $ \citep{Pohlhausen1921}, to infinite HAM series that remain valid over the entire range of $0\le \eta <\infty $ \citep{Liao2010}.  They have also given rise to a multitude of elegant approximations, such as those devised by \citet{Bairstow1925}, \citet{Parlange1981}, \citet{Boyd2008}, \citet{Iacono2015}, and other dedicated researchers.

Among those endeavours, several studies have devoted themselves to the determination of the Blasius constant with increasing degrees of success. For example, \citet{Bairstow1925} and \citet{Goldstein1930} reported 0.474 and 0.470, while \citet{Falkner1936} and \citet{Howarth1938} obtained 0.470334 and 0.469600. With the advent of modern computers, \citet{Fazio1992} and \citet{Boyd1999} arrived at 12 and 17 digits, which were later superseded by \citet{Abbasbandy2011}, who achieved 21 digits of accuracy with their 0.469599988361013304509 value.  The record for most significant digits at the time of this writing is held by \citet{Varin2014,Varin2018}: he manages to secure 30 and then 100 digits while expressing $\ddot{f}(0)$ in the form of a convergent series of rational numbers to arbitrary order. In this study, the availability of the Blasius constant with a high degree of precision is taken into full consideration while seeking higher-order solution refinements.

\begin{table}
\begin{center}
	\footnotesize
	\begin{tabular}{cc|cccccc}
	\toprule
$ \eta=\dfrac{y\sqrt{U}}{\sqrt{2\nu x}}$ & $ \xi=\dfrac{\eta}{\eta_{\delta}} $ & $ f(\eta) $ & $ \dfrac{\dd f} {\dd \eta}=\Fbar $& $ \dfrac{\dd^2 f} {\dd \eta^2} =\ddot{f}$ & $ \dfrac{\dd^3 f} {\dd \eta^3}=\dddot{f} $ &
$\dfrac{\dd \Fbar} {\dd \xi}=\dfrac{a \ddot{f}} {\sqrt{2}} $ & $\dfrac{\dd^2 \Fbar} {\dd \xi^2}=\dfrac{a \dddot{f}} {\sqrt{2}} $\\
\midrule
		0.0  &  0.00000  &  0.00000   &  0.00000   &   0.46960   &   $-$0.00000   &   1.63040   &  $-$0.00000      \\
	\midrule                                                                                                
		0.2  &  0.05761  &  0.00939  &   0.09391   &   0.46931  &   $-$0.00441   &   1.62938   &  $-$0.05313      \\
		0.4  &  0.11521  &  0.03755  &   0.18761   &   0.46725  &   $-$0.01755   &   1.62225   &  $-$0.21149      \\
		0.6  &  0.17282  &  0.08439  &   0.28058   &   0.46173  &   $-$0.03896   &   1.60309   &  $-$0.46967      \\
		0.8  &  0.23042  &  0.14967  &   0.37196   &   0.45119  &   $-$0.06753   &   1.56648   &  $-$0.81403      \\
		1.0  &  0.28803  &  0.23299  &   0.46063   &   0.43438  &   $-$0.10121   &   1.50812   &  $-$1.21994      \\
		\\                                                                                                 
		1.2  &  0.34563  &  0.33366  &   0.54525   &   0.41057  &   $-$0.13699   &   1.42544   &  $-$1.65126      \\
		1.4  &  0.40324  &  0.45072  &   0.62439   &   0.37969  &   $-$0.17114   &   1.31825   &  $-$2.06288      \\
		1.6  &  0.46084  &  0.58296  &   0.69670   &   0.34249  &   $-$0.19966   &   1.18908   &  $-$2.40664      \\
		1.8  &  0.51845  &  0.72887  &   0.76106   &   0.30045  &   $-$0.21899   &   1.04311   &  $-$2.63966      \\
		2.0  &  0.57606  &  0.88680  &   0.81669   &   0.25567  &   $-$0.22673   &   0.88765   &  $-$2.73296      \\
		\\                                                                                                 
		2.2  &  0.63366  &  1.05495  &   0.86330   &   0.21058  &   $-$0.22215   &   0.73111   &  $-$2.67780      \\
		2.4  &  0.69127  &  1.23153  &   0.90107   &   0.16756  &   $-$0.20636   &   0.58175   &  $-$2.48741      \\
		2.6  &  0.74887  &  1.41482  &   0.93060   &   0.12861  &   $-$0.18196   &   0.44653   &  $-$2.19340      \\
		2.8  &  0.80648  &  1.60328  &   0.95288   &   0.09511  &   $-$0.15249   &   0.33022   &  $-$1.83816      \\
		3.0  &  0.86408  &  1.79557  &   0.96905   &   0.06771  &   $-$0.12158   &   0.23508   &  $-$1.46551      \\
		\\                                                                                                 
		3.2  &  0.92169  &  1.99058  &   0.98036   &   0.04637  &   $-$0.09230   &   0.16099   &  $-$1.11263      \\
		3.3  &  0.95049  &  2.08883   &  0.98456   &   0.03781 &  $-$0.07899     &   0.13129   &  $-$0.95213      \\                                                                    
		3.4  &  0.97929  &  2.18747  &   0.98797   &   0.03054  &   $-$0.06679   &   0.10601   &  $-$0.80515      \\
		\midrule                                                                 
		3.47\dag&  1.00000  &  2.25856  &   0.99000   &   0.02603 & $-$0.05878   &   0.09036   &  $-$0.70854\\
		\midrule
		3.5  &  1.00810  &  2.28641   &  0.99071   &   0.02441   &  $-$0.05582   &   0.08477   &  $-$0.67288      \\
		3.6  &  1.03690  &  2.38559  &   0.99289   &   0.01933  &   $-$0.04611   &   0.06711   &  $-$0.55582      \\
		3.8  &  1.09451  &  2.58450  &   0.99594   &   0.01176  &   $-$0.03039   &   0.04082   &  $-$0.36633      \\
		4.0  &  1.15211  &  2.78389  &   0.99777   &   0.00687  &   $-$0.01914   &   0.02387   &  $-$0.23067      \\
%
		\bottomrule
		
	\end{tabular}
\caption{Numerical solution of the Blasius equation using both the similarity variable $\eta$ and the boundary-layer coordinate, $\xi=\eta/\eta_\delta=\eta\sqrt{2}/a$, where $\eta _{\delta} =3.471886880405967.\mathrm{\dagger}$}
	\label{Table_Blasius_solution}
	\end{center}
\end{table}

\subsection{Alignment of differential and integral predictions}\label{Confluence of differential and integral predictions}
\noindent In helping to establish the connection between the Blasius solution and the ongoing analysis of the KP approach, a well-resolved numerical solution of the Blasius equation is furnished in \tblref{Table_Blasius_solution}. This is performed using the continuous Taylor series method to compute several characteristic parameters with sufficient accuracy in all digits displayed. Interestingly, the first numerical solutions of the Blasius equation were generated manually by \citet{Toepfer1912} and, more precisely, by \citet{Howarth1938} and then \citet{Cortell2005}. \stblref{Table_Blasius_solution} catalogues values of the characteristic stream function $f$,  axial velocity $\dot{f}$, and shear stress $\ddot{f}$.  Other properties follow systematically. For example, since $\dot{f}=0.99$ at $\eta _{\delta } \approx 3.47188688$, we are able to deduce the 99\% boundary-layer thickness using 
\begin{equation}\label{Eq_disturbance_thickness}
	 \delta \equiv \delta _{99\% } \approx 3.47189\sqrt{\dfrac{2\nu x}{U} } \quad     \text{or}    \quad a = \dfrac{\delta }{x} \sqrt{Re_{x} } \approx 4.9099895 
\end{equation}
Note that the computed value for $a=\sqrt{2} \eta_{\delta } \approx 4.91$ is about 1.8\% lower than the traditional value of `5.0' computed manually \citep{Toepfer1912} and adopted still in various textbooks \citep{Batchelor2000,Fox2004,White2006,Fox2015,Anderson2017,Schlichting2017}. As for the wall shear stress and momentum thickness, they can be readily evaluated from
\begin{equation}\label{Eq_wall_shear_stress}
\tau _{w} =\dfrac{\mu U\ddot{f}(0)}{\sqrt{2\nu x/U} }\quad {\rm where} \quad \ddot{f}(0)=\int _{0}^{\infty }\dot{f}\left(1-\dot{f}\right)\dd\eta \approx  0.46959999 
\end{equation}
and so
\begin{equation}\label{Eq_skin_friction_coefficient}
C_{f} =\dfrac{2\tau _{w} }{\rho U^{2} } =\dfrac{{\it \theta }}{x} = \dfrac{b}{\sqrt{Re_{x} } } \quad    {\rm where} \quad     b=d= \frac{{\it \theta }}{x} \sqrt{Re_{x} } \approx 0.66411467
\end{equation}
\noindent Our refined Blasius calculations agree to varying degrees of accuracy with the coefficients associated with the integral approach; these have been specified earlier in \cref{Eq_a_constant,Eq_Cf_constant,Eq_ab2s,Eq_displacement_momentum_thicknesses}, particularly, where they are estimated using one of the velocity profiles in \cref{Eq_Pohlhausen,Eq_Schlichting,Eq_Majdalani-Xuan}.   Despite their basic agreement with the Blasius model, however, all piecewise approximations considered so far can be seen to suffer from two basic limitations. Firstly, they all end abruptly at $y=\delta$. Secondly, those considered up to the fourth polynomial order end with a value of $u=U$ instead of $u=0.99U$, thus slightly deviating from Prandtl's 99\% defining threshold at the boundary-layer edge.  To overcome these inconsistencies, we will turn our attention momentarily away from the analysis of piecewise approximations. Instead, we will proceed by leveraging the insight gained thus far to develop uniformly valid, continuously analytic profiles that mimic the behaviour of the Blasius solution more robustly and expansively, i.e., both in the viscous region and the far field.

\subsection{Procedure to determine continuous velocity profiles}

\noindent Based on several studies that seek to determine uniformly valid approximations of the Blasius equation \citep{Boyd1982,Boyd1999,Liao1999b,Parand2009}, one may infer that an exponential basis function is appropriate to pursue. This choice is further guided by the slow decay of the Blasius solution in the far field.  On this note, we proceed by adopting a function of the type, $F(\xi)=1-\exp[-Y(\xi)]$, where the structure of $Y(\xi)\rightarrow \infty$ as $\xi\rightarrow \infty$ is yet to be determined. 

Then based on our assessment of the piecewise solutions in \cref{Direct evaluation of properties using piecewise velocity approximations}--\cref{Extension to higher-order polynomial approximations}, we can infer that any refined formulation must seek to satisfy an amended form of Pohlhausen's requirements in \cref{Eq_noslip,Eq_smoothedge_BC,Eq_wall_momentum_balance_BC,Eq_edge_shear_BC}, i.e., by imposing realistic conditions that do not unnecessarily restrain the solution or force it to stray away from the Blasius distribution.  

This modification can be realized, first, by replacing the zero shear stress condition at $y=\delta$ by the actual velocity gradient at the wall, which plays a pivotal role in prescribing the model's alignment with the Blasius form (being representative of the wall vorticity).  Second, we insist on securing $u=0.99U$ at $y=\delta$, in conformance with Prandtl's strict definition of the boundary-layer thickness. Our amended conditions on $F(\xi)$ become: 
\begin{equation}\label{Eq_BC_Majdalani-Xuan}
F(0)=0, \quad F^{\prime} (0)=\sbar,\quad F^{\prime\prime}(0)=0, \quad F(1) =0.99  
\end{equation}

\noindent Note that we have eliminated Pohlhausen's fifth condition, $F^{\prime \prime}(1)=0$, in favour of a fixed gradient at the wall that equates to the Blasius constant, $\sbar$. We have also omitted Pohlhausen's fourth condition, which translates into $F^{\prime}(1)\approx 0.0904$.  Although additional constraints can be imposed, we find this assortment to be sufficient at this order and consistent with the Blasius model requirements. In fact, by writing $Y(\xi)$ as a cubic polynomial, we can systematically seek its coefficients based on \cref{Eq_BC_Majdalani-Xuan}. Using a polynomial basis for $Y(\xi)$, we can put
\begin{equation}\label{Eq_expo_series_expansion}
Y^{(N)}(\xi)=\sum_{n=0}^{N}{Y_n\xi^n};N=3
\end{equation}
A side benefit of this expansion is that the far-field condition, $F(\infty)\rightarrow 1$, becomes self-satisfied by the decaying exponential, granted that the highest-order coefficient $Y_N$ is positive. 
The remaining effort is straightforward and rather interesting. Firstly, the no-slip condition, $F(0)=0$, returns $1 - \exp(-Y_0)=0$, or $Y_0=0$.  Secondly, the fixed gradient, $F^{\prime} (0)=\sbar$, yields $Y_1=\sbar$.  Thirdly, the vanishing momentum balance at the wall, $F^{\prime\prime}(0)=0$, requires $Y_2=\sbar^2/2$. 
 We are now one constant away from securing our objective. The last trailing constant may be determined from Prandtl's formal definition of the boundary-layer cutoff point, namely, $F(1) =0.99$. By writing $1-\exp[-({\sbar}+\tfrac{1}{2}{\sbar}^2+Y_3)]=0.99$, we retrieve $Y_3=1.00937{\sbar}\approx {\sbar}$. Given the fortuitous nature of this outcome, we arrive at a surprisingly simple expression,
\begin{equation}\label{Eq_Majdalani-Xuan-Decaying-Expo}
F(\xi)=1-\exp[-{\sbar} \xi (1+\tfrac{1}{2}{\sbar}\xi+{\xi}^2)]
\end{equation}
Evidently, with a cubic polynomial representation for $Y(\xi)$, there is no freedom left to accommodate Pohlhausen's adjusted fourth condition, $F^{\prime}(1)\approx 0.0904$, on the asymptotically diminishing velocity gradient at the boundary-layer edge. Nonetheless, it is gratifying to see that \cref{Eq_Majdalani-Xuan-Decaying-Expo} still agrees with the corresponding Blasius gradient within $3.2\%$.

\subsection{Comparison to other continuous solutions}
\noindent In addition to \cref{Eq_Majdalani-Xuan-Decaying-Expo} and the piecewise velocity approximations detailed in \cref{Direct evaluation of properties using piecewise velocity approximations}, several continuous models have been developed with the goal of capturing the Blasius characteristics explored here.  Of those, three analytic solutions for $u/U$, which continue to hold past the $99\%$ disturbance thickness, will be considered and compared, after expressing them in terms of $\xi$.  These are:
 \begin{equation}\label{Eq_three_continuous_profiles}
F(\xi) =\left\{\begin{array}{ll} 
\tanh (\bar{s}\xi )    &\text{\citet{Yun2010}}  \\ 
\left[\tanh \left(\bar{s}\xi \right)^{5/3}\right]^{3/5}   &\text{\citet{Savas2012}}    \\
{\rm erf}(1.59261\xi )    &\text{\citet{Moeini2017}} 
\end{array}\right.     
\end{equation}
The foregoing approximations have been obtained using insight into the Blasius solution, rationalization, curve-fitting, or a combination thereof. 

To start, the ability of \cref{Eq_Majdalani-Xuan-Decaying-Expo} to reproduce the Blasius solution is confirmed in \tblref{Table_endpoint_properties_continuous_profiles}, where several values of $F(\xi)$ and its derivatives are evaluated and compared at both ends of the viscous layer. Compared to the other continuous profiles considered here, its predictions of $F'(0)$, $F(1)$, $F'(1)$, and $F''(1)$ appear to be the closest to the Blasius values except in the case of $F''(1)\approx-0.729$, where the $-0.722$ curvature in the error function of \citet{Moeini2017} deviates from the Blasius value of $-0.709$ by $1.8\%$, whereas \cref{Eq_Majdalani-Xuan-Decaying-Expo} differs by $2.9\%$.  Moreover, its prediction of $F'''(0)$ seems to be less precise than that of \citet{Savas2012}.

To be more specific, we are able to, firstly, corroborate the negative $F''(1)$ curvature of all velocity profiles, consistently with the $-0.709$ Blasius value at the edge of the viscous layer. The role of this negative curvature and its impact on Pohlhausen's quartic polynomial are discussed extensively in \cref{Paradoxical behavior of Pohlhausen's polynomial approximations}. Among the uniformly valid profiles, the closest curvature corresponds to Yun's hyperbolic tangent, while the farthest stems from Sava\c{s}'s hyperbolic tangent of fractional order.  Secondly, we readily confirm that all models satisfy the no-slip and momentum balance requirements at the wall, where $F(0)=F''(0)=0.$ Thirdly, both the decaying exponential and Sava\c{s} profiles return $u/U=99\% $ at $\xi =1$, whereas Moeini--Chamani's and Yun's return $97.6\%$ and $92.6\%$, respectively. As for the extraordinarily important velocity gradient at the wall, all three expressions by Sava\c{s}, Yun and \cref{Eq_Majdalani-Xuan-Decaying-Expo} restore the value of $F'(0)={\sbar}\approx 1.630$.  Finally, as far as matching the shear value (or velocity gradient) at the edge of the viscous layer, here too, both exponentially-decaying \cref{Eq_Majdalani-Xuan-Decaying-Expo} and Sava\c{s} profiles predict, respectively, $F'(1)\approx 0.093$ and $0.097$ in lieu of $0.0904$.  


In short, \cref{Eq_Majdalani-Xuan-Decaying-Expo} either matches identically the endpoint values of the Blasius distribution or deviates from them by less than $3.2\%$, up to the second derivative in $F$. Naturally, an even higher-order approximation can be devised by taking  $N=4,5,\dots,$ and securing additional conditions that are reflective of the Blasius solution, as performed in \cref{Extension to higher-order polynomial approximations}. When carefully constructed, all such solutions will have the intrinsic ability to approach unity as $\xi\rightarrow \infty$, thus satisfying the far-field condition identically. They will also possess the freedom to accommodate additional endpoint requirements with successive increases in the polynomial order.


\begin{table} 
		\begin{center}
		\begin{tabular}{l|c|c|c|c|c|c|c|c}
			\toprule
			$ \dfrac{u}{U} = F(\xi) $         & $ F(0) $ &  $ F'(0) $ & $ F''(0) $& $F^{\prime\prime\prime}(0)$ & $ F(1) $ & $ F'(1) $ & $ F''(1) $ & $ F^{\prime\prime\prime}(1) $ \\
			\midrule
			Blasius (1908)
			& 0		   & 1.63040       &0    & 0    &    0.99000 &$0.090357$&      $-0.70854$	& 4.4777\\
			\midrule
			$ 1-e^{-{\sbar}\xi(1+\tfrac{1}{2}{\sbar}\xi+\xi^2)} $ 
			& 0		   & 1.63040        &0     &$+$1.1145   &    0.98985     &     0.093211   &      $-0.72933$ & 4.4753	\\
			$ [\tanh\left(\bar{s}\xi\right)^{5/3}]^{3/5} $ 
			& 0		   & 1.63040        &0     &0   &    0.98698    &     0.097392   &      $-0.65884$ & 3.8398	\\
			$ \rm{erf}(1.59261\xi) $ 
			& 0        &  1.79707        &0     & $-9.1162$  &    0.97570     &     0.142239   &     $-0.72155$  & 2.9397    \\
			$ \tanh(\bar{s}\xi) $  
			& 0        & 1.63040        &  0    & $-8.6684$  &    0.92612     &     0.232014  &     $-0.70065$   & 1.9404    \\
			\midrule
			$1-\exp[-Y^{(4)}]$ \cref{Eq_expo_quartic_expansion}
			& 0        & 1.63040        & 0    & 0       &  0.99000  &  0.094266   & $-0.75123$  &4.6264 \\
			$1-\exp[-Y^{(5)}]$ \cref{Eq_expo_quintic_expansion}
			& 0        & 1.63040        & 0    & 0       &  0.99000  &   0.095424  &  $-0.76393$ &4.6676 \\
			$1-\exp[-Y^{(6)}]$  \cref{Eq_expo_sextic_expansion}
			& 0        & 1.63040        & 0    & 0       &  0.99000  &  0.090357   & $-0.70854$  &4.4678 \\
				$1-\exp[-Y^{(7)}]$ \cref{Eq_expo_septic_expansion}
			& 0		   & 1.63040       &0    & 0    &    0.99000 &       0.090357&      $-0.70854$	& 4.4777\\
			\bottomrule
		\end{tabular}
	\end{center}
	\caption{Endpoint properties of several continuous velocity profiles and their computed Blasius values. All profiles tend to unity as $\xi\rightarrow \infty.$ }
	\label{Table_endpoint_properties_continuous_profiles}
\end{table}

\begin{figure}
	\subfloat[]{\includegraphics*[width=0.49\textwidth]{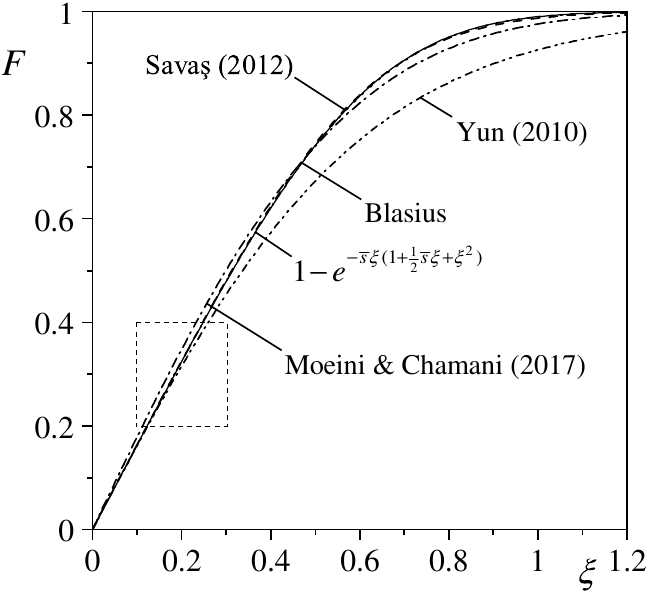}\label{Fig_Comparison_continuous_profiles_a}}
	\subfloat[]{\includegraphics*[width=0.49\textwidth]{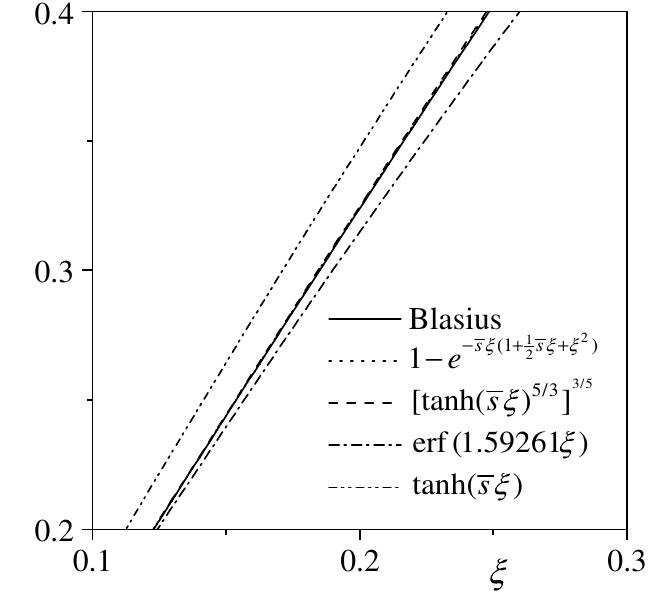}\label{Fig_Comparison_continuous_profiles_b}}
	\caption{Comparison of four continuous approximations for the Blasius solution according to \citet{Yun2010}, \citet{Savas2012}, \citet{Moeini2017}, and the systematically derived, decaying exponential function \cref{Eq_Majdalani-Xuan-Decaying-Expo}, which is imperceptible from the Blasius line. Results are shown (\textit{a}) across the boundary layer into the far field, and (\textit{b}) inside a designated quadrant where individual deviations from the Blasius curve are exaggerated.}
\label{Fig_Comparison_continuous_profiles}
\end{figure}

For further clarity, the four compact profiles in question are compared to the Blasius velocity function $\Fbar=\dot{f}$ in \cref{Fig_Comparison_continuous_profiles_a} for $0\le \xi \le 1.2$ ($0\le \eta \le 4.1663$), thus illustrating the degree by which they imitate the expected behaviour both within the boundary layer and beyond.  Graphically, it may be seen that the decaying exponential solution \cref{Eq_Majdalani-Xuan-Decaying-Expo} (dotted) is indiscernible from the Blasius curve. This may be viewed as being significant, given the relative simplicity of this profile.  This is followed by the Sava\c{s} hyperbolic tangent of fractional order, which can be barely distinguished from the Blasius line, even in the magnified inset of \cref{Fig_Comparison_continuous_profiles_b}.  Conversely, Moeini-Chamani's error function and Yun's hyperbolic tangent are shown to deviate progressively, especially in the upper $0.5\le\xi \le1$ portion of the viscous layer.  In fact, a strong correlation may be seen to exist between the spatial agreement of a given flow approximation with the Blasius shape and its effectiveness at predicting the fundamental boundary-layer characteristics.  This is confirmed in \tblref{Table_coefficient_comparison_continuous_profiles}, where boundary-layer predictions by the continuous profiles are catalogued and contrasted to the computed Blasius values. Unlike the results corresponding to the piecewise profiles in \cref*{Table_coefficient_comparison}, all integral properties that are posted here, such as $\alpha$ and $\beta$, are computed over the entire domain of validity, i.e., $0\le \xi < \infty$. As usual, the percentage error with respect to the Blasius benchmark, which accompanies each individual estimate, such as $\alpha,\beta,H,(\delta /x)\sqrt{Re_{x} } ,C_{f} \sqrt{Re_{x} } ,$  $(\delta\raisebox{+.15ex}{*}/x)\sqrt{Re_{x} } $, and $ (v/U)_{\delta} \sqrt{Re_{x} }$ is posted directly under it.  For each profile, we also compute and display the overall $L_{2} =[\int _{0}^{1}(F-\Fbar)^{2} \dd\xi]^{1/2} $ error across the viscous domain, to accurately quantify the cumulative discrepancy accrued in each continuous profile.  As before, only the essential properties are tabulated because other related quantities can be deduced fairly straightforwardly, e.g., $d=b=(\theta /x)\sqrt{Re_{x} } =C_{f} \sqrt{Re_{x} } ,$ $C_{D} \sqrt{Re_{L} } =2d$, and $G_{\infty}=[(\delta\raisebox{+.15ex}{*}/x)\sqrt{Re_{x} }]/2$. The latter prescribes the maximum normal velocity, $ (v/U)_{\rm max} \sqrt{Re_{x} }$, which occurs as $\xi \rightarrow \infty.$

\begin{table}
	\small\addtolength{\tabcolsep}{-2pt}
	\begin{center}
		\begin{tabular}{l|c|c|c|c|c|c|c|c}
			\toprule
			$F(\xi)=\dfrac{\dd f}{\dd \eta}=\dfrac{u}{U} $ & $ \alpha=\dfrac{\delta\raisebox{+.15ex}{*}}{\delta}$ & $ \beta=\dfrac{\theta}{\delta} $ & $ H=\dfrac{\alpha}{\beta} $ & $\dfrac{\delta}{x}\sqrt{Re_x}$ & $C_f\sqrt{Re_x}$ & $\dfrac{\delta\raisebox{+.15ex}{*}}{x}\sqrt{Re_x}$ & $  \left. \dfrac{v}{U} \right|_{\delta }\sqrt{Re_x} $ & $L_2 $ error\\
			\midrule
			Blasius &
			0.35047 &  0.13526 &  2.59110 &  4.90999  &  0.66411 &  1.72079 & 0.83340  &   0\\
			\midrule
			$ 1-e^{-{\sbar}\xi(1+\tfrac{1}{2}{\sbar}\xi+\xi^2)} $ &
			0.351  & 0.136  & 2.586  & 4.904 &  0.665  & 1.720 & 0.833 & $6.4\times 10^{-4}$\\
			Cubic model, \cref{Eq_Majdalani-Xuan-Decaying-Expo}&
			0.06\% &  0.25\%  &  0.19\% &   0.12\%  &   0.13\% &   0.06\%  & 0.10\%\\
			\midrule
			$ \left[\tanh\left(\bar{s}\xi\right)^{5/3}\right]^{3/5} $&
			0.351  & 0.136  & 2.574  & 4.891 &  0.667  & 1.716  & 0.824 &  $2.0\times 10^{-3}$\\
			Sava\c{s} (2012)&
			0.12\% &  0.79\%  &  0.67\% &   0.40\%  &   0.40\% &   0.27\% & 1.07\%\\	
			\midrule
			$ \rm{erf}(1.59261\xi) $ &
			0.354  & 0.147  & 2.414  & 4.949 &  0.726  & 1.753 & 0.800 &$1.6\times 10^{-2}$\\
			{Moeini-Chamani (2017)}&
			1.08\% &  8.49\%  &  6.83\% &   0.80\%  &   9.35\% &   1.89\% & 4.02\% \\
			\midrule
			$ \tanh(\bar{s}\xi) $ &
			0.425  & 0.188  & 2.259  & 4.163 &  0.783  & 1.770 & 0.805 &$5.6\times 10^{-2}$ \\
			Yun (2010) &
			21.30\% & 39.14\%  & 12.82\% &  15.23\%  &  17.96\% &   2.84\% & 3.45\% \\
			\midrule
			\midrule
			$1-\exp[-Y^{(4)}]$&  0.352  &  0.136  &  2.587  &  4.896   &  0.666   &  1.723   &  0.836 & $1.7\times 10^{-3}$\\
			Quartic model, \cref{Eq_expo_quartic_expansion}& 0.40\%  & 0.57\%  & 0.17\% & 0.29\%  & 0.29\%  & 0.12\%   & 0.31\% \\
			\midrule
			$1-\exp[{-Y^{(5)}}]$&  0.352  &  0.136  &  2.586  &  4.892   &  0.667   &  1.723   &  0.837 & $2.2\times 10^{-3}$\\
			Quintic model, \cref{Eq_expo_quintic_expansion}	& 0.53\%  & 0.74\%  & 0.22\% & 0.37\%  & 0.37\%  & 0.16\%   & 0.44\%\\
			\midrule
			$1-\exp[-Y^{(6)}]$&  0.350  &  0.135  &  2.591  &  4.911   &  0.664   &  1.720   &  0.832  &  $2.5\times 10^{-4}$\\
			Sextic model, \cref{Eq_expo_sextic_expansion}	& 0.06\%  & 0.05\%  & 0.01\% & 0.02\%  & 0.02\%  & 0.03\%   & 0.18\% \\
			\midrule
			$1-\exp[-Y^{(7)}]$&  0.350  &  0.135  &  2.591  &  4.910   &  0.664   &  1.721   &  0.832  & $1.1\times 10^{-4}$\\
			Septic model, \cref{Eq_expo_septic_expansion}& 0.02\%  & 0.02\%  & 0.01\% & 0.01\%  & 0.01\%  & 0.02\%   & 0.15\%\\
			\bottomrule
			
		\end{tabular}
	\end{center}
	\caption{ Boundary-layer predictions from several uniformly valid, continuous profiles with their errors relative to the computed Blasius values. }
	\label{Table_coefficient_comparison_continuous_profiles}
\end{table}

Starting with the $L_{2} $ error, we compute 0.056 and 0.016 for the Yun and Moeini--Chamani profiles, respectively.  These values are consistent with the level of spatial alignment between their curves and the Blasius distribution in \cref{Fig_Comparison_continuous_profiles}.  They are also reflected by their relative variations in estimating the boundary-layer properties, which range from 2.84\% to 39.14\% in Yun's case and from 0.8\% to 9.35\% in Moeini--Chamani's. 

In hindsight, these levels are comparable to the effectiveness of the Pohlhausen polynomial profiles, with Moeini--Chamani's results resembling those of Pohlhausen's quadratic polynomial in \eqref{Eq_Pohlhausen}.  In contrast, the overall $L_{2} $ disparity falls to appreciably low levels of $2.0\times 10^{-3} $ and $6.4\times 10^{-4} $ for the Sava\c{s} profile and the decaying exponential relation in \cref{Eq_Majdalani-Xuan-Decaying-Expo}, respectively.  These full orders of magnitude reductions may be viewed as being quite favourable because the corresponding percentage errors in estimating the principal boundary-layer properties suddenly decrease to virtually insignificant levels that do not exceed 0.79\% for the Sava\c{s} profile and 0.25\% for \cref{Eq_Majdalani-Xuan-Decaying-Expo}. These low levels can very well explain the reason why the decaying exponential form is imperceptible from the Blasius solution in \cref{Fig_Comparison_continuous_profiles}, including the graphically enlarged inset in \cref{Fig_Comparison_continuous_profiles_b}. In fact, recalling that the Blasius equation itself is accompanied by a small truncation error that depends on the size of the flow Reynolds number, and given that the Blasius model is only valid in the laminar range, the decaying exponential may be viewed as a simple, well-behaved, analytical solution of the Blasius equation in the $0\le \xi <\infty $ range. From an engineering perspective, however, the rationally-optimized quartic polynomial, $(5\xi -3\xi ^{3} +\xi ^{4} )/3$, remains, perhaps, among the simplest polynomial profiles that can be conveniently integrated to produce closed-form expressions for the main boundary-layer properties, albeit confined to the viscous region only.  For the remaining class of continuous profiles, numerical evaluation is generally required to determine their integral properties.

\subsection{Extension to higher-order decaying exponential solutions}
\label{Extension to higher-order decaying exponential solutions}
\noindent Using \cref{Eq_expo_series_expansion} and the cubic polynomial argument resulting in \cref{Eq_Majdalani-Xuan-Decaying-Expo}, it is possible to construct higher-order polynomial arguments that capture more precisely the conditions specified in \cref{Eq_Eight_BCs_for_polynomials}. To illustrate this process, we begin with $N=4$ in \cref{Eq_expo_series_expansion} and then apply five essential conditions, viz.
\begin{equation}\label{Eq_Five_BCs_for_Y4}  
\begin{gathered}
F(0)=0 \quad F'(0)={\sbar} \quad  F''(0)=0 \quad F'''(0)=0\quad \text{and}\quad F(1)=0.99\\
\end{gathered}
\end{equation} 
After some algebra, a quartic polynomial for the exponential argument is obtained in the form of
\begin{equation}\label{Eq_expo_quartic_expansion}
\begin{split}
Y^{(4)}(\xi)&\approx{{\sbar}\xi} \left[  X^{(c)}(\xi) +  \left(1.23244 {\sbar}-\tfrac{1}{3}{\sbar}^2-1\right)\xi^3\right]\approx Y^{(c)}(\xi) + 0.201033\xi^4
\end{split}
\end{equation}
where
\begin{equation}\label{Eq_common_quadratic_X}
\left\{\begin{array}{ll} 
	Y^{(c)}(\xi)\equiv& {\sbar}\xi X(\xi) \approx  1.6304 \xi+1.3291 \xi ^2+1.44464 \xi ^3 \\
	X^{(c)}(\xi)\equiv& 1 +\tfrac{1}{2}{\sbar} \xi +\tfrac{1 }{3}{\sbar}^2 \xi ^2\approx  1 +0.815199 \xi + 0.886066 \xi^2 \\
\end{array}\right.     
\end{equation}
Interestingly, due to the commonly shared physical requirements, the quadratic form of $X^{(c)}$ and its corresponding cubic expansion $Y^{(c)}$ will continue to recur at increasing orders. 

In seeking a higher-order model with $N=5$, the same procedure may be repeated with the addition of $F'''(1)\approx 4.4777041$ to \cref{Eq_Five_BCs_for_Y4}. The resulting quintic solution becomes
\begin{equation}\label{Eq_expo_quintic_expansion}
\begin{split}
Y^{(5)}(\xi)&\approx{{\sbar}\xi} \left[  X^{(c)} + 0.0523044 \xi ^3 +  0.0709987 \xi^4   \right]\approx Y^{(c)}+0.0852769 \xi ^4 +  0.115756 \xi ^5
\end{split}
\end{equation}
Note that only expansions with positive coefficients multiplying their highest-order terms are permitted to ensure that $Y(\infty)\rightarrow\infty$ and, as such, $F(\infty)\rightarrow 1$.  For this reason, the construction of a sextic polynomial requires exchanging the $F'''(1)$ condition with both $F'(1)\approx 0.090357643$ and $F''(1)\approx-0.70853762$, in addition to the five essential conditions in \cref{Eq_Five_BCs_for_Y4}. The strict application of all seven constraints enables us to extract a sextic solution of the form
\begin{equation}\label{Eq_expo_sextic_expansion}
\begin{split}
Y^{(6)}(\xi)&\approx{{\sbar}\xi} \left[  X^{(c)} +0.418226 \xi ^3 -0.350098 \xi ^4+ 0.0551749 \xi ^5  \right]\\
&\approx Y^{(c)}+ 0.681875 \xi^4 - 0.570799 \xi^5 + 0.0899571 \xi^6
\end{split}
\end{equation}

Finally, by imposing all eight conditions in \cref{Eq_Eight_BCs_for_polynomials}, a highly accurate approximation can be achieved, namely, 
\begin{equation}\label{Eq_expo_septic_expansion}
\begin{split}
Y^{(7)}(\xi)&\approx{{\sbar}\xi} \left[  X^{(c)} +0.335592 \xi^3 - 0.106739 \xi^4 - 0.183641 \xi^5 + 0.078091 \xi^6 \right]   \\
&\approx Y^{(c)} +0.547148 \xi ^4 -0.174026 \xi ^5  -0.299408 \xi ^6 +  0.127319 \xi ^7
\end{split}
\end{equation}



\noindent The validity of these higher-order solutions is showcased in \cref{Table_endpoint_properties_continuous_profiles}, where they are seen to mirror the behaviour of the Blasius solution and its derivatives at the endpoints of the viscous domain. Therein, it may be ascertained that additional endpoint conditions are secured with each successive increase in the polynomial order, with the septic model satisfying all eight requirements identically. 

The overall accuracy of these models is also affirmed in \cref{Table_coefficient_comparison_continuous_profiles}, where their ability to recover the characteristic boundary-layer properties is systematically tested. In this vein, it can be seen that the relative errors associated with both sextic and septic solutions are very small, being of $O(10^{-4})$, including their $L_2$ values.  For example, with an $L_2$ error of $1.08\times 10^{-4}$, the uniformly valid septic model appears to outperform most profiles in its class.  As for the quartic and quintic models, their errors  of $O(10^{-3})$ exceed those of their lower-order analogue in \cref{Eq_Majdalani-Xuan-Decaying-Expo}.  In fact, it may be speculated that the additional nonlinearities that accompany both quartic and quintic models in \cref{Eq_expo_quartic_expansion,Eq_expo_quintic_expansion} do not warrant their further pursuit, as they do not seem to offer tangible advantages over their lower-order expansion. Conversely, the simple three-term decaying exponential in \cref{Eq_Majdalani-Xuan-Decaying-Expo} stands as one of the most compact and accurate models in its class, being capable of surpassing its higher-order analogues in both overall precision and prediction of Blasius-related properties until the sixth order is achieved, albeit at twice the algebraic cost and number of nonlinear terms. Note that none of the profiles beyond \cref{Eq_Majdalani-Xuan-Decaying-Expo} appear in \cref{Fig_Comparison_continuous_profiles} as they become graphically indiscernible from the Blasius solution.

\section{Conclusion}

In this study, we revisit a classic paradigm in fluid mechanics, namely, the  K\'arm\'an-Pohlhausen (KP) momentum-integral approach, which has proven itself, time and time again, to be of tremendous academic and research value, especially in the analysis of viscous and thermal boundary layers.  Our work is carried out under the auspices of four overarching themes that aim at achieving four seemingly distinct but effectively joint objectives.  

The first aspiration, which has formed the initial motivation for this work, is the need to explain the paradoxical performance of Pohlhausen's high-order polynomials when used in concert with the momentum-integral approach. As such, our first objective has focused on pinpointing the technical factors that have caused Pohlhausen's quartic polynomial to accrue wider deviations during the evaluation of boundary-layer properties than its corresponding cubic and quadratic analogues. This perplexing behaviour, we have found, is attributable to three specific boundary conditions, conceived by Pohlhausen in 1921, with the ability to compel the solution to stray from its expected outcomes.  

Retrospectively, these conditions can be ranked in decreasing penalty levels. The first, which consists of a vanishing curvature at the boundary layer edge, $F''(1)=0$, is shown to be inconsistent with the asymptotic behaviour of the exact solution due to Blasius. The latter predicts a negative curvature of $F''(1)\approx -0.709$. The second, less conspicuous discrepancy, is traceable to the vanishing shear requirement, $F'(1)=0$, at the edge of the boundary layer, where the exact solution returns $F'(1)\approx0.09$. As for the third disparity, it is associated with the $F(1)=1$ condition, which mildly disagrees with Prandtl's strict definition of a boundary layer, i.e., as comprising the vertical distance to where $F(1)=0.99$. Realizing that the last two requirements, no matter how different from their expected values, are actually observed by Pohlhausen's quadratic and cubic polynomials, the reduced accuracy of the quartic profile relative to its lower-order expressions is pinned rather categorically on the prematurely imposed curvature requirement.  Fortuitously, the systemic process of sorting and comparing Pohlhausen's fundamental assumptions to the proper values of the Blasius velocity function and its derivatives across the viscous domain enables us to identify a more compliant set of boundary conditions.  In fact, by imposing a slightly more realistic assortment of boundary conditions, another quartic expansion is arrived at, specifically $F(\xi)=(5\xi-3\xi^{3}+\xi^{4} )/3$, which is accompanied by an order of magnitude reduction in error relative to the Blasius benchmark.  This effort, and its attendant optimization analysis, evolves into a second objective; it also results in a systematic procedure to derive rationally-optimized velocity profiles at successive polynomial orders.

For example, while constructing an alternative quartic profile, we start by retaining four of Pohlhausen's classic requirements; these include no slip and a judicious momentum balance at the wall coupled with a smooth merging of the velocity and its gradient at the boundary-layer edge.  We deem it essential to relax the zero curvature requirement and replace it with an objective function that accounts for the cumulative error residual affecting the most influential parameters in prescribing the boundary-layer properties. Besides the velocity gradient at the wall, the two other parameters that we designate consist of the non-dimensional displacement and momentum thicknesses. By linking these two integral quantities to the polynomial coefficients representing the profile in question, the objective function at the fourth order is reduced to a sole function of the slope, ${s}.$ The latter captures the wall vorticity and, as such, the initial velocity gradient. Subsequently, by minimizing the residual with respect to ${s}$, the optimal initial trajectory $s\raisebox{+.15ex}{{*}}$ is found in a manner to produce the smallest variance from the Blasius results with a quartic profile. 

The optimization procedure in \cref{Procedure to develop increasingly accurate polynomials}, however, cannot be undertaken in isolation.  To make further headway, linking the boundary-layer properties resulting from the KP integral approach to the actual polynomial coefficients in a velocity profile becomes, by necessity, our third objective.  In the ongoing analysis, this requirement is met through the development of several reduced momentum-integral relations in \cref{Reduced momentum-integral relations} that facilitate the evaluation of boundary-layer characteristics for a given velocity approximation. In fact, the resulting effort is found to be equally relevant to our first objective by virtue of its ability to enhance the effectiveness of tracing the source of deviation in any given flow profile, such as Pohlhausen's quartic polynomial, to the functional form of the velocity under construction. In summary, several simple relations are provided at the conclusion of this effort, and these enable us to deduce the fundamental boundary-layer characteristics from three principal quantities: the initial velocity gradient at the wall and both displacement and momentum thicknesses in non-dimensional form,  $\{s,\alpha,\beta\}$. Properties that can be readily retrieved using \cref{Eq_a_constant,Eq_Cf_constant,Eq_displacement_momentum_thicknesses,Eq_ab2s,Eq_ab_product,Eq_drag,Eq_v_velocity_derivation1,Eq_v_velocity_derivation2,Eq_v_velocity_derivation3,Eq_v_velocity_max,Eq_v_velocity_solution,Eq_v_max_coefficient,Eq_vorticity,Eq_vorticity_nondimensional} include the actual disturbance, displacement, and momentum thicknesses, the skin friction and drag coefficients, the wall-normal velocity and, let us not forget, the vorticity. This explains why the effort to obtain the most precise set of $\{s,\alpha,\beta\}$, from which all other properties can be derived, ends up dominating the optimization procedure. 

In fact, as the optimization procedure is further extended to higher polynomial orders, we find it essential to retain only three physical requirements, particularly, those that are matched identically by the Blasius model, i.e., $F(0)=F''(0)=0$, and $F(1)=0.99$. The remaining conditions are added incrementally to our objective function with successive increases in the polynomial order and the corresponding number of undetermined coefficients. Logistically, by continuing to minimize the overall residual error, we actively ensure that the remaining boundary conditions, which are turned into objective constraints, are collectively secured, as closely as possible, for the purpose of reducing deviations in boundary-layer predictions. This enables us to arrive in \cref{Extension to higher-order polynomial approximations} at the most effective quintic, sextic, and septic polynomials, which are capable of predicting the various boundary-layer metrics with an essentially $0.00\%$ error relative to their Blasius values. We project, although we only show for the case of a quartic profile in \cref{Procedure to develop increasingly accurate polynomials,Improved effectiveness in thermal analysis}, that straightforward integration of an optimally-constructed polynomial in viscous and thermal KP-type analyses will result in an overall improved framework. Naturally, since the error entailed in the KP approach is driven by the accuracy of the similarity-preserving velocity and temperature profiles that are adopted, the optimal refinement of the latter is due to favourably affect the former.

However, although polynomial approximations for velocity and temperature profiles are ideally suited for implementation in a global KP-type setting, where they facilitate the explicit evaluation of integral properties in closed form, they remain limited in their applicability to the viscous layer only. While comparing our optimal polynomials to a highly-resolved Blasius solution, as necessitated by the above-stated objectives, we are reminded of their piecewise analytic forms, which end abruptly at $\xi=1$.  We also realize that extending their range of validity into the far field will allow them to mirror the Blasius distribution even as $\xi\rightarrow\infty$, thus making them more realistic.  Then given the insight that we have gained into the merits of applying different endpoint requirements, and recognizing that the Blasius far-field condition, $F(\infty)=1,$ prohibits the use of a polynomial representation, we proceed to explore other options, such as the use of a decaying exponential for a more appropriate basis function, albeit at the expense of adding nonlinearities.  The remaining endeavour, which focuses on pursuing uniformly valid solutions to the Blasius problem, becomes our fourth and final objective. Although many attempts have been made to solve the Blasius problem, most of them have been wholly numerical, or resulting in infinite series solutions. Although there exist several integro-differential solutions that are formidably accurate \citep{Bairstow1925,Parlange1981,Boyd1997,Boyd2008, Iacono2015}, the complexity entailed in evaluating them prompts us to defer their analysis to future work.

As we redirect our investigation into the possibility of constructing uniformly valid solutions in \cref{Development of a continuous solution for the Blasius problem}, we exploit our understanding of the Blasius problem, which is developed while seeking to fulfil the other three overarching objectives, to identify five essential requirements that must be secured by a prospective formulation.  These consist of $F(0)=F^{\prime\prime}(0)=0,$ $F^{\prime} (0)={\sbar}$, $F(1) =0.99$, and $F(\infty)=1$. The latter is identically satisfied by the basis function that we explore, namely, a decaying exponential, $F(\xi)=1-\exp[-Y(\xi)]$. By casting $Y(\xi)$ in the form of a positive polynomial in the far field, we are left with four constraints that can be applied to extract a surprisingly accurate and compact expression, $F(\xi)=1-\exp[-{\sbar}\xi (1+\tfrac{1}{2}{\sbar}\xi+{\xi}^2)]$. By comparing this continuously analytic profile, which incurs an $L_2$ error of $6.4 \times 10^{-4}$, to other models in its class, its viability is cemented as an equivalent solution to the Blasius problem in most practical applications. However, should greater precision be desired, higher-order polynomial arguments, including sextic and septic polynomials, that entail even smaller errors, are also provided. To the authors' knowledge, the septic relation \cref{Eq_expo_septic_expansion}, which incurs an $L_2$ error of $1.08\times 10^{-4}$, may be representative of one of the most precise and uniformly valid analytical solutions of the Blasius problem using a relatively simple expression for the velocity itself, and not one of its derivatives.      

In closing, let us remark that this study, which was initially pursued in the context of streamlining the presentation of the KP momentum-integral analysis in a classroom setting, along with its connection to the Blasius problem, seems to have unravelled several useful findings.  Given the appreciable academic interest in the KP approach, as well as the Blasius equation, we hope that the various explanations provided here, including the procedure to improve the accuracy of the KP method, or to solve the Blasius problem using decaying exponential functions, could be further explored.

With the advent of an optimal quartic solution that complements Pohlhausen's cubic and quartic formulations, 
new challenges arise: The widely used $u=U(2\xi -2\xi ^{3} +\xi ^{4})$ model has been incorporated into countless viscous and thermal studies which, it stands to reason, ought to be revisited using the more precise quartic model.  This challenge can be daunting when taking into account that, in the absence of simpler alternatives over the course of a century, Pohlhausen's quartic polynomial has become the staple of analytic approximations characterizing the KP approach. In a sense, it has formed the backbone of numerous laminar-flow solutions of viscous and thermal boundary layers, both with and without pressure gradients \citep{Khan2005,Khan2006}. Evidently, its reduced accuracy has prompted several researchers to seek better alternatives, such as the methods of \citet{Walz1941} and \citet{Thwaites1949}. However, knowing that the accuracy of the KP approach can be markedly improved when used in conjunction with more refined velocity approximations, it is hoped that its applicability can be extended to problems with variable pressure gradients and freestream velocities, where other methods have superseded its use. Other curve-fit techniques, such as the method of Thwaites, are successful at providing global predictions, albeit with no information about the actual velocity that is driving the solution from within the boundary layer. In contrast, the KP approach retains the advantage of associating any of its solutions to tangible velocity and temperature fields.

For this reason, we hope that this study, which increases our repertoire of boundary-layer approximations and solutions to the Blasius problem, will open up additional areas of research endeavour. Specifically, we hope that our findings will encourage future investigations into seeking extensions of the KP approach, perhaps through the use of optimally-constructed velocity approximations, to other compressible, viscous and thermal flow applications.

\pdfbookmark[0]{Acknowledgments}{aknl}
\section*{Acknowledgements}
This work is supported partly by the National Science Foundation, through Grant No. 1761675, and partly by the Hugh and Loeda Francis Chair of Excellence, Department of Aerospace Engineering, Auburn University. The authors are deeply indebted to Dr. Martin J. Chiaverini, Director of Propulsion 
Systems at 	Sierra Nevada Corporation, for numerous technical exchanges and for his unwavering support of our flow field investigations.
\section*{Declaration of Interests}
The authors report no conflict of interest.

\bibliography{Majdalani_Xuan_KP_momentum_integral_approach}

\begin{thebibliography}{82}
\expandafter\ifx\csname natexlab\endcsname\relax\def\natexlab#1{#1}\fi
\def\au#1{#1} \def\ed#1{#1} \def\yr#1{#1}\def\at#1{#1}\def\jt#1{\textit{#1}}
  \def\bt#1{#1}\def\bvol#1{\textbf{#1}} \def\vol#1{#1} \def\pg#1{#1}
  \def\publ#1{#1}\def\arxiv#1{#1}\def\org#1{#1}\def\st#1{\textit{#1}}

\bibitem[Abbasbandy \& Bervillier(2011)]{Abbasbandy2011}
{\sc \au{Abbasbandy, S.} \& \au{Bervillier, C.}} \yr{2011}  \at{Analytic
  continuation of {T}aylor series and the boundary value problems of some
  nonlinear ordinary differential equations}.  \jt{Applied Mathematics and
  Computation}  \bvol{218}~(5),  \pg{2178--2199}.

\bibitem[Ahmad \& Al-Barakati(2009)]{Ahmad2009}
{\sc \au{Ahmad, F.} \& \au{Al-Barakati, W.~H.}} \yr{2009}  \at{An approximate
  analytic solution of the {B}lasius problem}.  \jt{Communications in Nonlinear
  Science and Numerical Simulation}  \bvol{14}~(4),  \pg{1021--1024}.

\bibitem[Anderson(2017)]{Anderson2017}
{\sc \au{Anderson, J.~D.}} \yr{2017} {\em Fundamentals of Aerodynamics\/}, 6th
  edn.  \publ{Princeton, NJ: McGraw-Hill}.

\bibitem[Andersson(1988)]{Andersson1988}
{\sc \au{Andersson, H.~I.}} \yr{1988}  \at{The {N}akayama-{K}oyama approach to
  laminar forced convection heat transfer to power-law fluids}.
  \jt{International Journal of Heat and Fluid Flow}  \bvol{9}~(3),  \pg{343 --
  346}.

\bibitem[Bairstow(1925)]{Bairstow1925}
{\sc \au{Bairstow, L.}} \yr{1925}  \at{Skin friction}.  \jt{The Journal of the
  Royal Aeronautical Society}  \bvol{29}~(169),  \pg{3--23}.

\bibitem[Batchelor(2000)]{Batchelor2000}
{\sc \au{Batchelor, G.~K.}} \yr{2000} {\em {An Introduction to Fluid
  Dynamics}\/}.  \publ{Cambridge: Cambridge University Press}, 67021953
  B67-17981 75/- by G.K. Batchelor. 24 plates, diagrs. 24 cm. Bibliography: p.
  604-608.

\bibitem[Beneitez {\em et~al.\/}(2019)Beneitez, Duguet, Schlatter \&
  Henningson]{Beneitez2019}
{\sc \au{Beneitez, M.}, \au{Duguet, Y.}, \au{Schlatter, P.} \& \au{Henningson,
  D.~S.}} \yr{2019}  \at{Edge tracking in spatially developing boundary layer
  flows}.  \jt{Journal of Fluid Mechanics}  \bvol{881},  \pg{164–181}.

\bibitem[Bizzell \& Slattery(1962)]{Bizzell1962}
{\sc \au{Bizzell, G.~D.} \& \au{Slattery, J.~C.}} \yr{1962}
  \at{Non-{N}ewtonian boundary-layer flow}.  \jt{Chemical Engineering Science}
  \bvol{17}~(10),  \pg{777--782}.

\bibitem[Blasius(1908)]{Blasius1908}
{\sc \au{Blasius, H.}} \yr{1908}  \at{Grenzschichten in {F}l\"ussigkeiten mit
  kleiner {R}eibung}.  \jt{Journal of Applied Mathematics and Mechanics (ZAMM)}
   \bvol{56},  \pg{1--37}.

\bibitem[Bloor \& Ingham(1977)]{Bloor1977}
{\sc \au{Bloor, M. I.~G.} \& \au{Ingham, D.~B.}} \yr{1977}  \at{On the use of a
  {P}ohlhausen method in three dimensional boundary layers}.  \jt{Journal of
  Applied Mathematics and Physics (ZAMP)}  \bvol{28},  \pg{289--299}.

\bibitem[Boyd(1982)]{Boyd1982}
{\sc \au{Boyd, J.~P.}} \yr{1982}  \at{The optimization of convergence for
  {C}hebyshev polynomial methods in an unbounded domain}.  \jt{Journal of
  Computational Physics}  \bvol{45}~(1),  \pg{43--79}.

\bibitem[Boyd(1997)]{Boyd1997}
{\sc \au{Boyd, J.~P.}} \yr{1997}  \at{Pad\'e approximant algorithm for solving
  nonlinear ordinary differential equation boundary value problems on an
  unbounded domain}.  \jt{Computers in Physics}  \bvol{11}~(3),  \pg{299--303}.

\bibitem[Boyd(1999)]{Boyd1999}
{\sc \au{Boyd, J.~P.}} \yr{1999}  \at{The {B}lasius function in the complex
  plane}.  \jt{Experimental Mathematics}  \bvol{8}~(4),  \pg{381--394}.

\bibitem[Boyd(2008)]{Boyd2008}
{\sc \au{Boyd, J.~P.}} \yr{2008}  \at{The {B}lasius function: Computations
  before computers, the value of tricks, undergraduate projects, and open
  research problems}.  \jt{SIAM Review}  \bvol{50}~(4),  \pg{791--804}.

\bibitem[Brynjell-Rahkola {\em et~al.\/}(2019)Brynjell-Rahkola, Hanifi \&
  Henningson]{Brynjell2019}
{\sc \au{Brynjell-Rahkola, M.}, \au{Hanifi, A.} \& \au{Henningson, D.~S.}}
  \yr{2019}  \at{On the stability of a {B}lasius boundary layer subject to
  localised suction}.  \jt{Journal of Fluid Mechanics}  \bvol{871},
  \pg{717–741}.

\bibitem[Bujurke \& Jagadeeswar(1992)]{Bujurke1992}
{\sc \au{Bujurke, N.~M.} \& \au{Jagadeeswar, M.}} \yr{1992}  \at{Momentum
  integral method in the analysis of taper-flat slider bearing with
  second-order fluid}.  \jt{Journal of Applied Mathematics and Mechanics
  (ZAMM)}  \bvol{72},  \pg{225--228}.

\bibitem[Callegari \& Nachman(1978)]{Callegari1978}
{\sc \au{Callegari, A.} \& \au{Nachman, A.}} \yr{1978}  \at{Some singular,
  nonlinear differential equations arising in boundary layer theory}.
  \jt{Journal of Mathematical Analysis and Applications}  \bvol{64}~(1),
  \pg{96--105}.

\bibitem[Callegari \& Friedman(1968)]{Callegari1968}
{\sc \au{Callegari, A.~J.} \& \au{Friedman, M.~B.}} \yr{1968}  \at{An
  analytical solution of a nonlinear, singular boundary value problem in the
  theory of viscous fluids}.  \jt{Journal of Mathematical Analysis and
  Applications}  \bvol{21}~(3),  \pg{510--529}.

\bibitem[Cebeci \& Cousteix(1998)]{Cebeci1998}
{\sc \au{Cebeci, T.} \& \au{Cousteix, J.}} \yr{1998} {\em Modeling and
  Computation of Boundary-Layer Flows\/}.  \publ{Long Beach, Calif: Horizons}.

\bibitem[Cortell(2005)]{Cortell2005}
{\sc \au{Cortell, R.}} \yr{2005}  \at{Numerical solutions of the classical
  {B}lasius flat-plate problem}.  \jt{Applied Mathematics and Computation}
  \bvol{170}~(1),  \pg{706--710}.

\bibitem[Ebaid \& Al-Armani(2013)]{Ebaid2013}
{\sc \au{Ebaid, A.} \& \au{Al-Armani, N.}} \yr{2013}  \at{A new approach for a
  class of the {B}lasius problem via a transformation and {A}domian’s
  method}.  \jt{Abstract and Applied Analysis}  \bvol{2013},  \pg{8}.

\bibitem[Falkner(1936)]{Falkner1936}
{\sc \au{Falkner, V.~M.}} \yr{1936}  \at{A method of numerical solution of
  differential equations}.  \jt{Philosophical Magazine}  \bvol{21}~(141),
  \pg{624--640}.

\bibitem[Fang {\em et~al.\/}(2008)Fang, Liang \& Lee]{Fang2008}
{\sc \au{Fang, T.}, \au{Liang, W.} \& \au{Lee, C.~F.}} \yr{2008}  \at{A new
  solution branch for the {B}lasius equation—a shrinking sheet problem}.
  \jt{Computers and Mathematics with Applications}  \bvol{56}~(12),
  \pg{3088--3095}.

\bibitem[Fazio(1992)]{Fazio1992}
{\sc \au{Fazio, R.}} \yr{1992}  \at{The {B}lasius problem formulated as a
  free-boundary value-problem}.  \jt{Acta Mechanica}  \bvol{95},  \pg{1--7}.

\bibitem[Fox {\em et~al.\/}(2004)Fox, McDonald \& Pritchard]{Fox2004}
{\sc \au{Fox, R.~W.}, \au{McDonald, A.~T.} \& \au{Pritchard, P.~J.}} \yr{2004}
  {\em Introduction to Fluid Mechanics\/}, 6th edn.  \publ{Hoboken, NJ: Wiley}.

\bibitem[Goldstein(1930)]{Goldstein1930}
{\sc \au{Goldstein, S.}} \yr{1930}  \at{Concerning some solutions of the
  boundary layer equations in hydrodynamics}.  \jt{Mathematical Proceedings of
  the Cambridge Philosophical Society}  \bvol{26}~(1),  \pg{1--30}.

\bibitem[Hack \& Moin(2017)]{Hack2017}
{\sc \au{Hack, M. J.~P.} \& \au{Moin, P.}} \yr{2017}  \at{Algebraic
  disturbance growth by interaction of {O}rr and lift-up mechanisms}.
  \jt{Journal of Fluid Mechanics}  \bvol{829},  \pg{112–126}.

\bibitem[Hewitt \& Duck(2018)]{Hewitt2018}
{\sc \au{Hewitt, R.~E.} \& \au{Duck, P.~W.}} \yr{2018}  \at{Localised streak
  solutions for a {B}lasius boundary layer}.  \jt{Journal of Fluid Mechanics}
  \bvol{849},  \pg{885–901}.

\bibitem[Howarth(1938)]{Howarth1938}
{\sc \au{Howarth, L.}} \yr{1938}  \at{On the solution of the laminar boundary
  layer equations}.  \jt{Proceedings of the Royal Society of London, Series A}
  \bvol{164}~(919),  \pg{547--579}.

\bibitem[Huang \& Garc\'ia(1998)]{Huang1998}
{\sc \au{Huang, X.} \& \au{Garc\'ia, M.~H.}} \yr{1998}  \at{A
  {H}erschel–{B}ulkley model for mud flow down a slope}.  \jt{Journal of
  Fluid Mechanics}  \bvol{374},  \pg{305–333}.

\bibitem[Iacono \& Boyd(2015)]{Iacono2015}
{\sc \au{Iacono, R.} \& \au{Boyd, J.~P.}} \yr{2015}  \at{Simple analytic
  approximations for the {B}lasius problem}.  \jt{Physica D: Nonlinear
  Phenomena}  \bvol{310},  \pg{72--78}.

\bibitem[Jain \& Kumar(1972)]{Jain1972}
{\sc \au{Jain, A.~C.} \& \au{Kumar, A.}} \yr{1972}  \at{Hypersonic rarefied
  flow past an insulated flat plate with suction/injection}.  \jt{International
  Journal of Heat and Mass Transfer}  \bvol{15},  \pg{2401--2407}.

\bibitem[Khan {\em et~al.\/}(2005)Khan, Culham \& Yovanovich]{Khan2005}
{\sc \au{Khan, W.~A.}, \au{Culham, J.~R.} \& \au{Yovanovich, M.~M.}} \yr{2005}
  \at{Fluid flow around and heat transfer from an infinite circular cylinder}.
  \jt{Journal of Heat Transfer}  \bvol{127},  \pg{785--790}.

\bibitem[Khan {\em et~al.\/}(2006)Khan, Culham \& Yovanovich]{Khan2006}
{\sc \au{Khan, W.~A.}, \au{Culham, J.~R.} \& \au{Yovanovich, M.~M.}} \yr{2006}
  \at{Fluid flow and heat transfer in power-law fluids across circular
  cylinders: Analytical study}.  \jt{Journal of Heat Transfer}  \bvol{128}~(9),
   \pg{870--878}.

\bibitem[Levin {\em et~al.\/}(2005)Levin, Chernoray, L{{\"{o}}}fdahl \&
  Henningson]{Levin2005}
{\sc \au{Levin, O.}, \au{Chernoray, V.~G.}, \au{L{{\"{o}}}fdahl, L.} \&
  \au{Henningson, D.~S.}} \yr{2005}  \at{A study of the {B}lasius wall jet}.
  \jt{Journal of Fluid Mechanics}  \bvol{539},  \pg{313–347}.

\bibitem[Liao \& Campo(2002)]{Liao2002}
{\sc \au{Liao, S.} \& \au{Campo, A.}} \yr{2002}  \at{Analytic solutions of the
  temperature distribution in {B}lasius viscous flow problems}.  \jt{Journal of
  Fluid Mechanics}  \bvol{453},  \pg{411–425}.

\bibitem[Liao(1997)]{Liao1997}
{\sc \au{Liao, S.~J.}} \yr{1997}  \at{A kind of approximate solution technique
  which does not depend upon small parameters{ — II.} {A}n application in
  fluid mechanics}.  \jt{International Journal of Non-Linear Mechanics}
  \bvol{32}~(5),  \pg{815--822}.

\bibitem[Liao(1999{\natexlab{{\em a\/}}})]{Liao1999a}
{\sc \au{Liao, S.~J.}} \yr{1999{\natexlab{{\em a\/}}}}  \at{An explicit,
  totally analytic approximate solution for {B}lasius’ viscous flow
  problems}.  \jt{International Journal of Non-Linear Mechanics}
  \bvol{34}~(4),  \pg{759--778}.

\bibitem[Liao(1999{\natexlab{{\em b\/}}})]{Liao1999b}
{\sc \au{Liao, S.~J.}} \yr{1999{\natexlab{{\em b\/}}}}  \at{A uniformly valid
  analytic solution of two-dimensional viscous flow over a semi-infinite flat
  plate}.  \jt{Journal of Fluid Mechanics}  \bvol{385},  \pg{101--128}.

\bibitem[Liao(2010)]{Liao2010}
{\sc \au{Liao, S.~J.}} \yr{2010}  \at{An optimal homotopy-analysis approach for
  strongly nonlinear differential equations}.  \jt{Communications in Nonlinear
  Science and Numerical Simulation}  \bvol{15}~(8),  \pg{2003--2016}.

\bibitem[Mehta \& Sood(1994)]{Mehta1994}
{\sc \au{Mehta, K.~N.} \& \au{Sood, Shobha}} \yr{1994}  \at{Free convection
  about axisymmetric bodies immersed in inhomogeneous porous medium}.
  \jt{International Journal of Engineering Science}  \bvol{32},  \pg{945--953}.

\bibitem[Moeini \& Chamani(2017)]{Moeini2017}
{\sc \au{Moeini, M.} \& \au{Chamani, M.~R.}} \yr{2017}  \at{New perspectives on
  the laminar boundary layer physics in a polarized pressure field with
  temperature gradient: An analytical approximation to {B}lasius equation}.
  \jt{Journal of Applied Fluid Mechanics}  \bvol{10}~(4),  \pg{1071--1077}.

\bibitem[Mohanty {\em et~al.\/}(1977)Mohanty, Raghavachar \&
  Nanda]{Mohanty1977}
{\sc \au{Mohanty, A.K.}, \au{Raghavachar, T.S.} \& \au{Nanda, R.S.}} \yr{1977}
  \at{Heat transfer from rotating short radial blades}.  \jt{International
  Journal of Heat and Mass Transfer}  \bvol{20},  \pg{1417--1425}.

\bibitem[Mustapha \& Mohand(2003)]{Amaouche2003}
{\sc \au{Mustapha, A.} \& \au{Mohand, K.}} \yr{2003}  \at{On {C}auchy
  conditions for asymmetric mixed convection boundary layer flows}.
  \jt{International Journal of Thermal Sciences}  \bvol{42},  \pg{621--630}.

\bibitem[Myers(2010)]{Meyers2010}
{\sc \au{Myers, T.~G.}} \yr{2010}  \at{An approximate solution method for
  boundary layer flow of a power law fluidover a flat plate}.
  \jt{International Journal of Heat and Mass Transfer}  \bvol{53},
  \pg{2337–2346}.

\bibitem[Nakayama \& Koyama(1988)]{Nakayama1988}
{\sc \au{Nakayama, A.} \& \au{Koyama, H}} \yr{1988}  \at{An analysis for
  friction and heat transfer characteristics of power-law non-{N}ewtonian fluid
  flows past bodies of arbitrary geometrical configuration}.  \jt{W\"{a}rme-
  und Stoff\"{u}bertragung}  \bvol{22},  \pg{29–36}.

\bibitem[Nakayama \& Pop(1989)]{Nakayama1989}
{\sc \au{Nakayama, A.} \& \au{Pop, I.}} \yr{1989}  \at{Free convection over a
  nonisothermal body in a porous medium with viscous dissipation}.
  \jt{International Communications in Heat and Mass Transfer}  \bvol{16},
  \pg{173--180}.

\bibitem[Nakayama {\em et~al.\/}(1986)Nakayama, Shenoy \& Koyama]{Nakayama1986}
{\sc \au{Nakayama, A.}, \au{Shenoy, A.~V.} \& \au{Koyama, H}} \yr{1986}  \at{An
  analysis for forced convection heat transfer from external surfaces to
  non-{N}ewtonian fluids}.  \jt{W\"{a}rme- und Stoff\"{u}bertragung}
  \bvol{20},  \pg{219–227}.

\bibitem[Ng {\em et~al.\/}(2017)Ng, Ooi, Lohse \& Chung]{Ng2017}
{\sc \au{Ng, C.~S.}, \au{Ooi, A.}, \au{Lohse, D.} \& \au{Chung, D.}} \yr{2017}
  \at{Changes in the boundary-layer structure at the edge of the ultimate
  regime in vertical natural convection}.  \jt{Journal of Fluid Mechanics}
  \bvol{825},  \pg{550–572}.

\bibitem[Oleinik \& Samokhin(1999)]{Oleinik1999}
{\sc \au{Oleinik, O.~A.} \& \au{Samokhin, V.~N.}} \yr{1999} {\em Mathematical
  Models in Boundary Layer Theory\/}.  \publ{London: CRC Press}.

\bibitem[Parand {\em et~al.\/}(2009)Parand, Dehghan \& Pirkhedri]{Parand2009}
{\sc \au{Parand, K.}, \au{Dehghan, M.} \& \au{Pirkhedri, A.}} \yr{2009}
  \at{Sinc-collocation method for solving the {B}lasius equation}.  \jt{Physics
  Letters A}  \bvol{373}~(44),  \pg{4060--4065}.

\bibitem[Parlange {\em et~al.\/}(1981)Parlange, Braddock \&
  Sander]{Parlange1981}
{\sc \au{Parlange, J.~Y.}, \au{Braddock, R.~D.} \& \au{Sander, G.}} \yr{1981}
  \at{Analytical approximations to the solution of the {B}lasius equation}.
  \jt{Acta Mechanica}  \bvol{38}~(1),  \pg{119--125}.

\bibitem[de~Paula {\em et~al.\/}(2017)de~Paula, W\"urz, Mendon\c{c}a \&
  Medeiros]{Depaula2017}
{\sc \au{de~Paula, I.~B.}, \au{W\"urz, W.}, \au{Mendon\c{c}a, M.~T.} \&
  \au{Medeiros, M. A.~F.}} \yr{2017}  \at{Interaction of instability waves and
  a three-dimensional roughness element in a boundary layer}.  \jt{Journal of
  Fluid Mechanics}  \bvol{824},  \pg{624–660}.

\bibitem[Peker {\em et~al.\/}(2011)Peker, Karao\u{g}lu \&
  Oturan\c{c}]{Peker2011}
{\sc \au{Peker, H.~A.}, \au{Karao\u{g}lu, O.} \& \au{Oturan\c{c}, G.}}
  \yr{2011}  \at{The differential transformation method and {P}ade approximant
  for a form of {B}lasius equation}.  \jt{Mathematical and Computational
  Applications}  \bvol{16}~(2),  \pg{507--513}.

\bibitem[Pohlhausen(1921)]{Pohlhausen1921}
{\sc \au{Pohlhausen, K.}} \yr{1921}  \at{{Z}ur n\"{a}herungsweisen
  {I}ntegration der {D}ifferentialgleichung der laminaren {G}renzschicht}.
  \jt{Journal of Applied Mathematics and Mechanics (ZAMM)}  \bvol{1}~(4),
  \pg{252--290}.

\bibitem[Prandtl(1904)]{Prandtl1904}
{\sc \au{Prandtl, L.}} \yr{1904}  \at{{\"{U}}ber {F}l\"{u}ssigkeitsbewegungen
  bei sehr kleiner {R}eibung}.  \jt{Verhandlungen des III. Internationalen
  Mathematiker Kongresses}  \pg{p. 484–491}.

\bibitem[Pritchard \& Mitchell(2015)]{Fox2015}
{\sc \au{Pritchard, P.~J.} \& \au{Mitchell, John~W.}} \yr{2015} {\em Fox and
  McDonald's Introduction to Fluid Mechanics\/}, 9th edn.  \publ{Hoboken, NJ:
  Wiley}.

\bibitem[Punnis(1956{\natexlab{{\em a\/}}})]{Punnis1956a}
{\sc \au{Punnis, B.}} \yr{1956{\natexlab{{\em a\/}}}}  \at{{Z}ur
  {D}ifferentialgleichung der {P}lattengrenzschicht von {B}lasius}.
  \jt{Journal of Applied Mathematics and Mechanics (ZAMM)}  \bvol{36}~(S1),
  \pg{S26--S26}.

\bibitem[Punnis(1956{\natexlab{{\em b\/}}})]{Punnis1956b}
{\sc \au{Punnis, B.}} \yr{1956{\natexlab{{\em b\/}}}}  \at{{Z}ur
  {D}ifferentialgleichung der {P}lattengrenzschicht von {B}lasius}.  \jt{Archiv
  der Mathematik}  \bvol{7}~(3),  \pg{165--171}.

\bibitem[Rao \& Arakeri(1998)]{Rao1998}
{\sc \au{Rao, A.} \& \au{Arakeri, J.~H.}} \yr{1998}  \at{Integral analysis
  applied to radial film flows}.  \jt{International Journal of Heat and Mass
  Transfer}  \bvol{41},  \pg{2757--2767}.

\bibitem[Rosenhead(1963)]{Rosenhead1963}
{\sc \au{Rosenhead, L.}} \yr{1963} {\em Laminar Boundary Layer\/}.
  \publ{London: Oxford Univ. Press}.

\bibitem[Sava\c{s}(2012)]{Savas2012}
{\sc \au{Sava\c{s}, \"{O}}} \yr{2012}  \at{An approximate compact analytical
  expression for the {B}lasius velocity profile}.  \jt{Communications in
  Nonlinear Science and Numerical Simulation}  \bvol{17}~(10),
  \pg{3772--3775}.

\bibitem[Schetz \& Bowersox(2011)]{Schetz2011}
{\sc \au{Schetz, J.~A.} \& \au{Bowersox, R. D.~W.}} \yr{2011}  \at{Boundary
  layer analysis}.  \jt{AIAA Education Series, 2nd ed., American Institute of
  Aeronautics and Astronautics}  \bvol{31}~(1),  \pg{257--260}.

\bibitem[Schlichting(1955)]{Schlichting1955}
{\sc \au{Schlichting, H.}} \yr{1955} {\em Boundary-Layer Theory\/}, 1st edn.
  \publ{New York: McGraw-Hill}.

\bibitem[Schlichting(1979)]{Schlichting1979}
{\sc \au{Schlichting, H.}} \yr{1979} {\em Boundary-Layer Theory\/}, 7th edn.
  \publ{New York: McGraw-Hill}.

\bibitem[Schlichting \& Gersten(2017)]{Schlichting2017}
{\sc \au{Schlichting, H.} \& \au{Gersten, K.}} \yr{2017} {\em Boundary-Layer
  Theory\/}, 9th edn.  \publ{Berlin: Springer-Verlag}.

\bibitem[Shenoy \& Nakayama(1986)]{Shenoy1986}
{\sc \au{Shenoy, A.~V.} \& \au{Nakayama, A.}} \yr{1986}  \at{Forced convection
  heat transfer from axisymmetric bodies to non-{N}ewtonian fluids}.  \jt{The
  Canadian Journal of Chemical Engineering}  \bvol{64}~(4),  \pg{680--686}.

\bibitem[Thayalan \& Hung(2013)]{Thayalan2013}
{\sc \au{Thayalan, N.} \& \au{Hung, Y.~M.}} \yr{2013}  \at{Momentum integral
  method for forced convection in thermal nonequilibrium power-law
  fluid-saturated porous media}.  \jt{Chemical Engineering Communications}
  \bvol{200}~(2),  \pg{269--288}.

\bibitem[Thwaites(1949)]{Thwaites1949}
{\sc \au{Thwaites, B.}} \yr{1949}  \at{Approximate calculation of the laminar
  boundary layer}.  \jt{Aeronautical Quarterly}  \bvol{1},  \pg{245–280}.

\bibitem[T\"{o}epfer(1912)]{Toepfer1912}
{\sc \au{T\"{o}epfer, K.}} \yr{1912}  \at{{B}emerkung zu dem {A}ufsatz von h.
  {B}lasius: {G}renzschichten in {F}l\"{u}ssigkeiten mit kleiner {R}eibung}.
  \jt{Journal of Applied Mathematics and Physics (ZAMP)}  \bvol{60},
  \pg{397--398}.

\bibitem[Tsang {\em et~al.\/}(2018)Tsang, Dalziel \& Vriend]{Tsang2018}
{\sc \au{Tsang, J. M.~F.}, \au{Dalziel, S.~B.} \& \au{Vriend, N.~M.}} \yr{2018}
   \at{Interaction between the {B}lasius boundary layer and a free surface}.
  \jt{Journal of Fluid Mechanics}  \bvol{839},  \pg{R1}.

\bibitem[Tsang {\em et~al.\/}(2019)Tsang, Dalziel \& Vriend]{Tsang2019}
{\sc \au{Tsang, J. M.~F.}, \au{Dalziel, S.~B.} \& \au{Vriend, N.~M.}} \yr{2019}
   \at{The granular {B}lasius problem}.  \jt{Journal of Fluid Mechanics}
  \bvol{872},  \pg{784–817}.

\bibitem[Tsigklifis \& Lucey(2017)]{Tsigklifis2017}
{\sc \au{Tsigklifis, K.} \& \au{Lucey, A.~D.}} \yr{2017}  \at{The interaction
  of {B}lasius boundary-layer flow with a compliant panel: global, local and
  transient analyses}.  \jt{Journal of Fluid Mechanics}  \bvol{827},
  \pg{155–193}.

\bibitem[Varin(2014)]{Varin2014}
{\sc \au{Varin, V.~P.}} \yr{2014}  \at{A solution of the {B}lasius problem}.
  \jt{Computational Mathematics and Mathematical Physics}  \bvol{54}~(6),
  \pg{1025--1036}.

\bibitem[Varin(2018)]{Varin2018}
{\sc \au{Varin, V.~P.}} \yr{2018}  \at{Asymptotic expansion of {C}rocco
  solution and the {B}lasius constant}.  \jt{Computational Mathematics and
  Mathematical Physics}  \bvol{58}~(4),  \pg{517--528}.

\bibitem[Von~K\'{a}rm\'{a}n(1921)]{Karman1921}
{\sc \au{Von~K\'{a}rm\'{a}n, T.}} \yr{1921}  \at{\"{U}ber laminare und
  turbulente {R}eibung}.  \jt{Journal of Applied Mathematics and Mechanics
  (ZAMM)}  \bvol{1}~(4),  \pg{233--252}.

\bibitem[Walz(1941)]{Walz1941}
{\sc \au{Walz, A.}} \yr{1941}  \at{Ein neuer {Ansatz} f\"{u}r das
  {Greschwindligkeitsprofil} der laminaren {R}eibungsschicht}.
  \jt{Lilienthal-Bericht}  \bvol{141},  \pg{8}.

\bibitem[Wang(2004)]{Wang2004}
{\sc \au{Wang, L.}} \yr{2004}  \at{A new algorithm for solving classical
  {B}lasius equation}.  \jt{Applied Mathematics and Computation}
  \bvol{157}~(1),  \pg{1 -- 9}.

\bibitem[Weyl(1942)]{Weyl1942}
{\sc \au{Weyl, H.}} \yr{1942}  \at{On the differential equations of the
  simplest boundary-layer problems}.  \jt{Annals of Mathematics}
  \bvol{43}~(2),  \pg{381--407}.

\bibitem[White(2006)]{White2006}
{\sc \au{White, F.~M.}} \yr{2006} {\em Viscous Fluid Flow\/}, 3rd edn.
  \publ{New York: McGraw-Hill}.

\bibitem[Yun(2010)]{Yun2010}
{\sc \au{Yun, B.~I.}} \yr{2010}  \at{Intuitive approach to the approximate
  analytical solution for the {B}lasius problem}.  \jt{Applied Mathematics and
  Computation}  \bvol{215}~(10),  \pg{3489--3494}.

\bibitem[Zhao {\em et~al.\/}(2017)Zhao, Lei \& Patterson]{Zhao2017}
{\sc \au{Zhao, Y.}, \au{Lei, C.} \& \au{Patterson, J.~C.}} \yr{2017}  \at{The
  {K}-type and {H}-type transitions of natural convection boundary layers}.
  \jt{Journal of Fluid Mechanics}  \bvol{824},  \pg{352–387}.

\end{thebibliography}
\end{document}